\PassOptionsToPackage{unicode}{hyperref}
\PassOptionsToPackage{hyphens}{url}
\PassOptionsToPackage{dvipsnames,svgnames,x11names}{xcolor}
\documentclass[
  12pt]{article}

\usepackage[utf8]{inputenc}
\usepackage[T1]{fontenc}
\usepackage{hyperref}
\usepackage{url}
\usepackage{microtype}
\usepackage{xcolor}
\usepackage{amsmath,amssymb,amsthm}
\usepackage{amsfonts}
\usepackage{nicefrac}
\usepackage{booktabs}
\usepackage{algorithm}
\usepackage{algorithmic}
\usepackage{graphicx}

\newtheorem{proposition}{Proposition}
\newcommand{\Ga}{\mathrm{Ga}}
\newcommand{\E}{\mathbb{E}}
\newcommand{\Var}{\mathrm{Var}}
\newcommand{\Cov}{\mathrm{Cov}}

\usepackage{amsmath,amssymb}
\usepackage{iftex}
\usepackage{textcomp} 
\usepackage{lmodern}
\IfFileExists{upquote.sty}{\usepackage{upquote}}{}
\IfFileExists{microtype.sty}{
  \usepackage[]{microtype}
  \UseMicrotypeSet[protrusion]{basicmath} 
}{}
\makeatletter
\@ifundefined{KOMAClassName}{
  \IfFileExists{parskip.sty}{%
    \usepackage{parskip}
  }{
    \setlength{\parindent}{0pt}
    \setlength{\parskip}{6pt plus 2pt minus 1pt}}
}{
  \KOMAoptions{parskip=half}}
\makeatother
\usepackage{xcolor}
\makeatletter
\ifx\paragraph\undefined\else
  \let\oldparagraph\paragraph
  \renewcommand{\paragraph}{
    \@ifstar
      \xxxParagraphStar
      \xxxParagraphNoStar
  }
  \newcommand{\xxxParagraphStar}[1]{\oldparagraph*{#1}\mbox{}}
  \newcommand{\xxxParagraphNoStar}[1]{\oldparagraph{#1}\mbox{}}
\fi
\ifx\subparagraph\undefined\else
  \let\oldsubparagraph\subparagraph
  \renewcommand{\subparagraph}{
    \@ifstar
      \xxxSubParagraphStar
      \xxxSubParagraphNoStar
  }
  \newcommand{\xxxSubParagraphStar}[1]{\oldsubparagraph*{#1}\mbox{}}
  \newcommand{\xxxSubParagraphNoStar}[1]{\oldsubparagraph{#1}\mbox{}}
\fi
\makeatother

\usepackage{longtable,booktabs,array}
\usepackage{calc} 
\usepackage{etoolbox}
\makeatletter
\patchcmd\longtable{\par}{\if@noskipsec\mbox{}\fi\par}{}{}
\makeatother
\IfFileExists{footnotehyper.sty}{\usepackage{footnotehyper}}{\usepackage{footnote}}
\makesavenoteenv{longtable}
\usepackage{graphicx}
\makeatletter
\def\maxwidth{\ifdim\Gin@nat@width>\linewidth\linewidth\else\Gin@nat@width\fi}
\def\maxheight{\ifdim\Gin@nat@height>\textheight\textheight\else\Gin@nat@height\fi}
\makeatother
\setkeys{Gin}{width=\maxwidth,height=\maxheight,keepaspectratio}
\makeatletter
\def\fps@figure{htbp}
\makeatother

\makeatletter
\@ifpackageloaded{caption}{}{\usepackage{caption}}
\AtBeginDocument{%
\ifdefined\contentsname
  \renewcommand*\contentsname{Table of contents}
\else
  \newcommand\contentsname{Table of contents}
\fi
\ifdefined\listfigurename
  \renewcommand*\listfigurename{List of Figures}
\else
  \newcommand\listfigurename{List of Figures}
\fi
\ifdefined\listtablename
  \renewcommand*\listtablename{List of Tables}
\else
  \newcommand\listtablename{List of Tables}
\fi
\ifdefined\figurename
  \renewcommand*\figurename{Figure}
\else
  \newcommand\figurename{Figure}
\fi
\ifdefined\tablename
  \renewcommand*\tablename{Table}
\else
  \newcommand\tablename{Table}
\fi
}
\@ifpackageloaded{float}{}{\usepackage{float}}
\floatstyle{ruled}
\@ifundefined{c@chapter}{\newfloat{codelisting}{h}{lop}}{\newfloat{codelisting}{h}{lop}[chapter]}
\floatname{codelisting}{Listing}

\makeatother
\makeatletter
\@ifpackageloaded{caption}{}{\usepackage{caption}}
\@ifpackageloaded{subcaption}{}{\usepackage{subcaption}}
\makeatother

\ifLuaTeX
  \usepackage{selnolig}  
\fi
\usepackage[]{natbib}
\usepackage{bookmark}

\IfFileExists{xurl.sty}{\usepackage{xurl}}{} 
\hypersetup{
  pdftitle={Title},
  pdfauthor={Author 1; Author 2},
  pdfkeywords={3 to 6 keywords, that do not appear in the title},
  colorlinks=true,
  linkcolor={blue},
  filecolor={Maroon},
  citecolor={Blue},
  urlcolor={Blue},
  pdfcreator={LaTeX via pandoc}}

\newcommand{\anon}{1}

\begin{document}

\def\spacingset#1{\renewcommand{\baselinestretch}%
{#1}\small\normalsize} \spacingset{1}


\if1\anon
{
  \title{\bf DISCCO: Distance-Based Bayesian Spatial Clustering of Complex Objects with Node-Frailty Centrality}
  \author{Srijato Bhattacharyya,
    Huiyan Sang, and
    Bani Mallick \\
    Department of Statistics, Texas A\&M University}
  \date{}
  \maketitle
} \fi

\if0\anon
{
  \bigskip
  \bigskip
  \bigskip
  \begin{center}
    {\LARGE\bf DISCCO: Distance-Based Bayesian Spatial Clustering of Complex Objects with Node-Frailty Centrality}
\end{center}
  \medskip
} \fi

\bigskip
\begin{abstract}
Clustering problems increasingly involve complex objects observed over space, such as distributions, matrices, functions, images, or multivariate data, for which a scientifically meaningful dissimilarity between objects is often easier to specify and computationally more tractable than an object-response likelihood model. We propose \textsc{DISCCO}, a Bayesian framework for clustering spatially indexed complex objects using only a pairwise distance matrix and a spatial adjacency graph, with broad applicability and minimal user modeling requirements. The model combines a hierarchical distance-based likelihood accounting for within-cluster compactness and between-cluster separation with a random spatial graph partition prior, ensuring that posterior clusters are spatially contiguous. A key model feature is a set of node-specific frailty parameters that induce dependence among overlapping within-cluster distances and provide posterior summaries of object-level centrality or peripherality within each inferred cluster. We develop a partially collapsed Markov chain Monte Carlo algorithm for posterior inference. Simulations with distribution- and matrix-valued responses show that the proposed spatial distance-clustering framework improves region recovery relative to existing distance-clustering methods, while the frailty layer provides interpretable centrality summaries. Real applications to Houston Census Block Group racial-composition distributions and Western US county-level cancer mortality matrices illustrate how the method recovers interpretable contiguous clusters and frailty-based centrality maps.
\end{abstract}

\noindent%
\vfill

\newpage
\spacingset{1.8} 

\section{Introduction}\label{sec-intro}

Modern spatial data often place a \emph{complex object} at each location rather than a scalar outcome: a transcriptomic profile, a function-valued curve, a multivariate signature, or a spatio-temporal sequence \citep{zhao2021spatial,allen2023bayesian,hu2023bayesian,zhang2023bayesian,mozdzen2022bayesian}. For such data, a scientifically meaningful pairwise dissimilarity is frequently more natural and more robust than a bespoke likelihood on the raw observation scale, which has fueled recent interest in Bayesian clustering methods defined directly on dissimilarities \citep{duan2021bayesian,natarajan2024cohesion}. In spatial problems, however, similarity alone is not enough: the objective is to discover \emph{contiguous} regions whose associated objects are alike, together with uncertainty quantification for those regions. Bridging that gap is the central challenge of this paper.

Bayesian clustering is built on a rich model-based foundation in which uncertainty over partitions is represented explicitly through random partition priors and hierarchical mixture constructions, including Dirichlet process and Pitman--Yor process formulations and their hierarchical extensions \citep{ferguson1973Bayesian,pitman1997two,griffiths2003hierarchical,teh2006hierarchical,blei2010nested,orbanz2008nonparametric,muller2015Bayesian}. A closely related line of work develops product-partition and generalized product-partition models, which place priors directly on partitions and provide a flexible basis for incorporating covariate information and structured dependence into clustering \citep{hartigan1990partition,barry1992product,quintana2003Bayesian,park2010Bayesian,muller2011product,quintana2022dependent}. In spatial settings, these ideas have led to a broad Bayesian literature on contiguous region discovery, including spatial product-partition models and related random partition formulations that encourage or enforce local coherence \citep{jo2015spatial,page2016spatial,page2019spatiotemporal}. More recently, graph-based approaches have made spatial contiguity explicit by modeling connected partitions directly, including spanning-tree partition methods, contiguous graph-partition priors, and constrained random graph partitions \citep{teixeira2019bayesian,luo2021bayesian,luo2021bast,bhattacharyya2025constrained}. Collectively, these methods provide elegant probabilistic tools for spatial clustering and uncertainty quantification, but they are typically formulated on the \emph{raw observation scale}, with application-specific likelihoods for scalar response models. As a result, even though the spatial partitioning machinery is well developed, these methods are not a natural fit when each location carries a complex object and the scientifically meaningful representation is a pairwise dissimilarity matrix rather than a generative likelihood for the raw data.

A natural response is to work directly with pairwise dissimilarities rather than with a generative model for the original observations. This perspective has motivated a growing literature on Bayesian clustering with dissimilarity data. Early approaches typically relied on latent Euclidean or model-based embeddings that treat the observed dissimilarities as indirect surrogates for an underlying representation \citep{oh2007model,morrissette2024parsimonious}, while more recent methods define the clustering model more directly on the pairwise distances themselves \citep{duan2021bayesian,natarajan2024cohesion}. These developments are especially appealing for high-dimensional, functional, image-valued, and other non-Euclidean data, where a meaningful dissimilarity may be available even when likelihood-based modeling on the original scale is difficult or unnatural. However, existing Bayesian distance-clustering methods remain fundamentally non-spatial: they do not exploit neighborhood structure, do not enforce contiguity of the recovered partition, and therefore do not directly address the region-discovery problems that arise in spatial applications \citep{duan2021bayesian,natarajan2024cohesion}. Moreover, their likelihood constructions are built from collections of pairwise distance contributions, but the dependence induced by overlapping distances is not modeled explicitly. In particular, distances such as $d_{ij}$ and $d_{ik}$ share the common unit $i$, and in many applications that shared object is expected to induce systematic local association. Accounting for this dependence is important when the distance matrix is modeled as a structured statistical object rather than as an unrelated collection of pairwise summaries. Moreover, the dependence induced by shared nodes need not be viewed solely as a nuisance feature. In many applications, it contains scientifically meaningful information about the role of individual locations within an inferred region: some locations may be highly representative of the within-cluster geometry, whereas others may be comparatively peripheral or transitional. A distance-based spatial clustering model should therefore both accommodate the dependence among overlapping dissimilarities and provide posterior inference on the resulting notion of object-level centrality. Existing Bayesian distance-clustering methods do not directly yield such node-specific centrality summaries from the observed distance matrix.

We address these problems by introducing \textsc{DISCCO}, a Bayesian method for contiguous clustering of spatially indexed complex objects from pairwise dissimilarities. The model embeds a distance-based likelihood for within-cluster compactness and between-cluster separation within a spanning-tree prior over connected graph partitions. Its distinguishing feature is a set of node-specific frailty parameters that enter all within-cluster dissimilarities involving a given location. These frailties induce dependence among overlapping distances and provide posterior estimates of object-level centrality within each spatial cluster. Thus, DISCCO recovers spatially coherent clusters while also identifying locations that are central or peripheral within their inferred clusters. \textit{Throughout, centrality refers to centrality in the within-cluster distance geometry, rather than graph-theoretic centrality in the spatial adjacency graph.} At the computational level, we develop a partially collapsed Markov chain Monte Carlo algorithm built on local tree-cut proposals, exact tree-refresh moves, and latent-effect updates. Once a scientifically meaningful distance matrix has been constructed, the same model and algorithm can be applied across different types of complex objects without developing an object-specific response likelihood. Conjugacy permits the cluster-level compactness and separation parameters to be integrated out analytically, allowing partition proposals to be evaluated through local updates of collapsed likelihood factors. In addition, an exact conditional update of the compatible spanning tree avoids approximate exploration of the tree space. By collapsing the cluster-level parameters while retaining the node frailties, the sampler explores connected partitions in a fixed-dimensional state space and yields posterior summaries for the partition, the number of clusters, and node-level centrality. 

The remainder of the paper is organized as follows: Section~\ref{sec:model} introduces the Bayesian model, Section~\ref{sec:cov} studies its key covariance properties, Section~\ref{sec:computation} describes posterior computation, and Sections~\ref{sec:simulation} and~\ref{sec:realdata} present simulation studies and real-data applications.

\section{Model}
\label{sec:model}

We seek a contiguous partition of a spatial domain into blocks that group locations carrying similar complex objects. Let $V=\{1,\ldots,n\}$ index the spatial units, let $G=(V,E_G)$ be a connected spatial adjacency graph describing their adjacency structure, and suppose that for each unordered pair of distinct units $(i,j)$ a positive distance $d_{ij}>0$ between the complex objects observed at those units is available. 
Let $D=(d_{ij})$ be the symmetric distance matrix with $d_{ij}=d_{ji}$ and $d_{ii}=0$, and let $d  = \{d_{ij}:1\le i<j\le n\}$ denote its upper-triangular off-diagonal entries. Our goal is to infer a contiguous partition $\rho=\{S_1,\ldots,S_K\}$ of $V$, where each $S_h$ induces a connected subgraph of $G$, which can be naturally interpreted as a spatial cluster.  


\noindent\textbf{Distance-based likelihood with node frailties.}
We define the likelihood directly on the distances, with separate model components for pairs within the same cluster and pairs in different clusters. Throughout, $\Ga(\alpha,\beta)$ denotes a Gamma distribution with shape $\alpha$ and rate $\beta$.
If units $i$ and $j$ belong to the same cluster $S_h$, we assume
\begin{equation}
d_{ij}\mid \lambda_h,w_i,w_j \stackrel{ind} \sim \Ga(\delta_w,\lambda_h w_i w_j), \qquad 
\lambda_h \stackrel{ind}\sim \Ga(a_\lambda,b_\lambda), \qquad 
w_i \stackrel{ind}\sim \Ga(\kappa,\kappa).
\label{eq:within_likelihood}
\end{equation}

Hence $\mathbb{E}[d_{ij}\mid \lambda_h,w_i,w_j]=\delta_w/(\lambda_h w_iw_j)$. The parameter $\lambda_h$ controls cluster-level compactness, with larger values corresponding to smaller expected within-cluster distances. The frailty $w_i$ is a unit-level effect shared across all within-cluster distances involving unit $i$; holding $\lambda_h$ fixed, larger $w_i$ decreases the expected distances from unit $i$ to other members of its region. The parameter $\lambda_h$ therefore modulates the overall scale of within-cluster distances in $S_h$, while $w_i$ locally adjusts all within-cluster distances involving unit $i$ in the same direction.

If $i\in S_h$ and $j\in S_\ell$ with $h\neq \ell$, we assume
\begin{equation}
d_{ij}\mid \theta_{h\ell} \stackrel{ind}\sim \Ga(\delta_b,\theta_{h\ell}), \qquad 
\theta_{h\ell} \stackrel{ind}\sim \Ga(a_\theta,b_\theta).
\label{eq:between-likelihood}
\end{equation}

Thus $\mathbb{E}[d_{ij}\mid \theta_{h\ell}]=\delta_b/\theta_{h\ell}$. The parameter $\theta_{h\ell}$ controls the scale of distances between regions $S_h$ and $S_\ell$. Thus the likelihood separates three sources of variation in the distance matrix: within-cluster compactness through $\lambda_h$, between-cluster separation through $\theta_{h\ell}$, and unit-level within-cluster departures through the frailties $w_i$.

The prior $w_i\sim \Ga(\kappa,\kappa)$ centers the frailties at one, with $\mathbb{E}[w_i]=1$ and $\Var(w_i)=1/\kappa$. Thus $\kappa$ controls the prior degree of unit-level heterogeneity: larger values shrink the $w_i$'s toward one, while smaller values allow greater variation across units. This normalization anchors the multiplicative scale of $\lambda_h w_iw_j$ and keeps $\lambda_h$ interpretable as the cluster-level compactness parameter. For singleton clusters, $w_i$ is informed only by its prior because no within-cluster distances are observed, so frailty-based centrality summaries are used only for non-singleton clusters. The conjugate Gamma specifications ensure positive rate parameters and yield the collapsed likelihood factors used in posterior computation. 

\noindent\textbf{Spatial partition prior.}
To enforce contiguity of the spatial clusters, we place a prior on partitions through a tree-cut representation. Let $T$ denote a spanning tree of the adjacency graph $G$, that is, a connected acyclic subgraph connecting all vertices in $V$. 
Let $B\subseteq E(T)$ be a set of tree edges. Since $T$ is a tree, deleting $|B|$ edges breaks it into exactly $K=|B|+1$ connected components, each of which is a connected subgraph of $G$ and therefore defines a spatially contiguous cluster with respect to $G$. We write the induced partition as $\rho(T,B)=\{S_1,\ldots,S_K\}$.

We place a prior on the tree-cut-based partition through
\[
\pi_0(T,B,K,\rho\mid G)=p(T\mid G)\,p(K)\,p(B\mid T,K)\,\mathbb{I}\{\rho=\rho(T,B)\}.
\]
Accordingly, the generative partition prior is defined through three ingredients: a prior on the spanning tree, a prior on the number of clusters, and a prior on the cut set conditional on $(T,K)$. In our implementation, $p(T\mid G)$ is taken to be the discrete uniform prior over all spanning trees of $G$, $p(K)\propto \eta^{K-1}$ is a truncated geometric prior on $\{1,\ldots,n\}$, and $p(B\mid T,K)$ is uniform over all subsets of $E(T)$ of size $K-1$, so that $p(B\mid T,K)=\binom{n-1}{K-1}^{-1}$ whenever $|B|=K-1$. Owing to treating the tree as random, this model supports all possible contiguous partitions with respect to the original spatial graph $G$, thus preserving partition richness. This specification is simple, interpretable, and computationally convenient, while still inducing substantial flexibility over connected partitions as shown in \citep{teixeira2019bayesian,luo2021bayesian,bhattacharyya2025constrained}. Moreover, because the partition is induced through edge deletions that generate arbitrary connected subgraphs, it can accommodate highly irregular and non-convex clusters. This makes the prior especially well-suited for spatial clustering, where one seeks contiguous regions of similar units without imposing restrictive shape assumptions. 

\noindent\textbf{Scope and applicability.}
The proposed framework applies whenever two ingredients are available: a connected spatial graph describing local adjacency and a scientifically meaningful positive dissimilarity matrix summarizing dissimilarity among the observed objects. It is therefore well suited to settings in which each location carries a complex response, including multivariate vectors, matrices, functions, images, networks, or other structured data for which likelihood-based modeling on the original scale may be difficult or unnatural. By treating the dissimilarity matrix as the primary data object while preserving spatial contiguity through the graph prior, the model provides a general Bayesian framework for region discovery in complex spatial data. 

\section{Centrality and Dependence in the Distance Matrix}
\label{sec:cov}

The node frailties $w_i$, introduced in Section~\ref{sec:model}, have two formal implications given a partition. First, within a non-singleton cluster, larger frailty values imply smaller expected within-cluster distances involving that unit, after accounting for the cluster-level compactness parameter. This yields a model-induced ordering of units by relative centrality in the within-cluster distance geometry. Second, because the same frailty appears in all within-cluster distances involving a given unit, marginalizing over the latent frailties induces stronger dependence between distances that share an endpoint than between disjoint within-cluster distances. This section formalizes both these properties.

To formalize the notion of frailty-based centrality, fix a non-singleton cluster $S_h$ with $n_h=|S_h|\ge 2$ and let $i\in S_h$. Write $d_{ij}$ for a random distance generated under the model. Motivated by statistical depth ideas that rank observations by centrality or inverse outlyingness relative to a data cloud \citep{vardi2000multivariate,zuo2000general}, define the frailty-adjusted remoteness of unit $i$ within cluster $h$ as $R_i^{(h)}=(n_h-1)^{-1}\sum_{j\in S_h\setminus\{i\}} w_jd_{ij}$. Since $\mathbb{E}[d_{ij}\mid \lambda_h,w,\rho]=\delta_w/(\lambda_hw_iw_j)$ for $i,j\in S_h$, it follows that $\mathbb{E}[R_i^{(h)}\mid \lambda_h,w,\rho]=\delta_w/(\lambda_hw_i)$. The weight $w_j$ in the expression of remoteness removes the contribution of the comparison unit from the conditional mean, so that the expected adjusted remoteness only reflects the role of unit $i$. The associated model-induced depth-like centrality is $\mathrm{Dep}_i^{(h)}=\{\mathbb{E}[R_i^{(h)}\mid \lambda_h,w,\rho]\}^{-1}$,  and hence $\mathrm{Dep}_i^{(h)}=(\lambda_h/\delta_w)w_i$. Therefore, for any $i,k\in S_h$, $w_i>w_k$ if and only if $\mathrm{Dep}_i^{(h)}>\mathrm{Dep}_k^{(h)}$. Thus, within a fixed non-singleton cluster, $w_i$ induces the same ordering as a model-based inverse-remoteness score, justifying posterior frailty summaries as a relative centrality score in the within-cluster distance geometry.

\noindent\textbf{Frailty-induced covariance structure.}
The same frailty construction also determines how distances co-vary after marginalizing over the latent rates and frailties. 
\begin{proposition}
\label{prop:frailty_covariance}
Assume $a_\lambda>2$, $a_\theta>2$, and $\kappa>2$, so that the relevant second moments exist.
Under the proposed likelihood, the following properties hold.
\begin{enumerate}
    \item If $i,j,k,\ell$ are distinct and belong to the same cluster, then within-cluster distances are positively correlated, with shared-node pairs more strongly associated than disjoint pairs: $\Cov(d_{ij},d_{ik})>\Cov(d_{ij},d_{k\ell})>0$.
    \item If $i,i'\in S_h$ and $j,j'\in S_\ell$ for two distinct clusters $S_h$ and $S_\ell$, then $\Cov(d_{ij},d_{i'j'})>0$.
    \item If two between-cluster distances are associated with different unordered pairs of clusters, then they are independent.
    \item A within-cluster distance and a between-cluster distance are independent.
\end{enumerate}
\end{proposition}

Proof of Proposition~\ref{prop:frailty_covariance} is provided in the Supplementary materials. These covariance statements clarify the roles of the latent variables $\lambda_h$ and $w_i$ in the distance model. Within a cluster, the common compactness parameter $\lambda_h$ creates positive marginal dependence among all within-cluster distances, including disjoint pairs such as $d_{ij}$ and $d_{k\ell}$. Node frailties add a second source of dependence: if two distances share an endpoint, such as $d_{ij}$ and $d_{ik}$, both involve the same latent variable $w_i$, producing additional covariance beyond the cluster-level effect. Consequently, shared-node within-cluster distances are more strongly correlated than disjoint within-cluster distances. Without the frailties, the model would retain dependence through $\lambda_h$, but it would not distinguish overlapping from disjoint within-cluster pairs.

Between clusters, dependence is induced only through the block-pair separation parameter. All distances connecting the same unordered pair of clusters $\{S_h,S_\ell\}$ share $\theta_{h\ell}$ and are therefore exchangeably positively correlated. Distances associated with different unordered block pairs are independent, and within-cluster distances are independent of between-cluster distances, because they depend on disjoint sets of latent variables. Thus the model uses $\lambda_h$ for cluster-level compactness, $\theta_{h\ell}$ for between-cluster separation, and $w_i$ for shared-node dependence and relative centrality summaries within clusters. Together, the inverse-remoteness notion and the covariance properties show that the parameters in the DISCCO likelihood model are highly interpretable.

\section{Posterior computation}
\label{sec:computation}

Posterior inference uses a partially collapsed Markov chain Monte Carlo sampler to iteratively draw posterior samples of the tree-cut partition state
$\Omega = (T,B,K,\rho)$ and node frailties $w=\{w_i\}$. 

We analytically integrate out the cluster-level compactness parameters $\lambda_h$ and the between-cluster separation parameters $\theta_{h\ell}$ to facilitate mixing, while retaining $w$ in the posterior draws so that frailty-based posterior summaries remain available for centrality interpretation. The resulting partially collapsed posterior is
\[
\pi(T,B,K,\rho,w\mid d,G)\propto \pi_0(T,B,K,\rho\mid G)p(w\mid\kappa)\prod_{h=1}^K W_w(S_h)\prod_{1\le h<\ell\le K}R(S_h,S_\ell).
\]
Here $p(w\mid\kappa)=\prod_{i=1}^n p(w_i\mid\kappa)$, and $W_w(S)$ and $R(S,S')$ are the collapsed within-cluster and between-cluster distance likelihood factors derived from \eqref{eq:within_likelihood} and \eqref{eq:between-likelihood}, respectively; their closed forms are given in the Supplementary materials.

\noindent\textbf{Tree-cut updates of $(B,K,\rho)$.}
Conditional on $w$ and $T$, we draw samples of the tree-cut state by three local Metropolis--Hastings moves. A split move adds an uncut edge of spanning tree $T$ to the edge cut set $B$, increasing the cluster number $K$ by one and dividing one component into two connected components. A merge move removes a cut edge from $B$, decreasing $K$ by one and joining two adjacent components. A cut-swap move keeps $K$ fixed by merging two adjacent components and then cutting a different edge in the resulting connected subtree. These moves preserve spatial contiguity by construction. Because a proposed move changes only the affected components, the collapsed likelihood ratio can be computed from local updates to the within-cluster factors $W_w(\cdot)$ and between-cluster factors $R(\cdot,\cdot)$. 
Let $\ell(\rho,w)=\sum_{h=1}^K\log W_w(S_h)+\sum_{1\le h<\ell\le K}\log R(S_h,S_\ell)$ denote the collapsed distance score. For a proposed tree-cut state $x'$, the Metropolis--Hastings ratio combines the prior ratio, the local change in $\ell(\rho,w)$, and the proposal ratio. The move probabilities and Metropolis--Hastings ratios are given in the Supplementary materials.

\noindent\textbf{Tree updates of $T$.}
We sample $T$ from its posterior conditional on the current $\rho$, a typically challenging task due to the large combinatorial tree space, so previous methods based on similar tree priors adopt an approximate sampler~\citep{teixeira2019bayesian}. We derive a closed-form full conditional and an exact sampler: we sample a uniform spanning tree on the quotient graph obtained by contracting each cluster, independently sample uniform spanning trees within clusters, and graft these pieces to obtain a compatible tree on $G$. Cut set $B$ is updated accordingly in this procedure. 
Implementation details are given in the Supplementary materials. This improves mixing by allowing subsequent split, merge, and cut-swap proposals to use different tree representations of the same partition. 

\noindent\textbf{Frailty updates of $w$.}
Suppose unit $i$ belongs to a cluster $S$. Since $w_i$ appears only in the within-cluster factor for its own block and in its prior, its full conditional depends only on the frailties and distances within $S$, denoted as $w_{-i,S}$. This allows an efficient Metropolis-within-Gibbs sampler, where conditional independence across blocks enables parallel updates, while parameters within each block are updated sequentially.
Let $n_S=|S|$, $m_S=\binom{n_S}{2}$, and $q_S=a_\lambda+\delta_w m_S$. For a non-singleton block, write
$D_S^{(w)}=\sum_{u<v:\,u,v\in S}w_uw_vd_{uv}=C_i+w_iH_i$,
where $C_i=\sum_{u<v:\,u,v\in S\setminus\{i\}}w_uw_vd_{uv}$ and $H_i=\sum_{j\in S\setminus\{i\}}w_jd_{ij}$. The full conditional of $w_i$ is then, up to proportionality,
\[
p(w_i\mid w_{-i,S},\rho,d)\propto w_i^{\kappa-1+\delta_w(n_S-1)}\exp(-\kappa w_i)(b_\lambda+C_i+w_iH_i)^{-q_S}, \qquad i \in S, w_i>0.
\]
For singleton blocks, the within-cluster likelihood does not involve $w_i$, so the update reduces to the prior. For non-singleton blocks, we update $\eta_i=\log w_i$ using a one-dimensional random-walk Metropolis algorithm.
Implementation details are given in the Supplementary materials.

Because DISCCO uses the full pairwise dissimilarity matrix, the current
implementation requires $O(n^2)$ storage. This quadratic cost is inherent to
retaining all $\binom{n}{2}$ dissimilarities and is not introduced by the
tree-cut prior. Importantly, the local partition updates described above avoid recomputing the full pairwise likelihood at each proposal. A complete frailty sweep can still have quadratic cost in the current implementation, although the updates are one-dimensional and
conditionally independent across blocks, allowing parallelization. 
For substantially larger graphs, the quadratic burden could be reduced by replacing the full pairwise likelihood with a sparse approximation based on a selected set of informative pairs, such as spatial neighbors or nearest neighbors under the observed dissimilarity, and by using graph coarsening to reduce the effective number of spatial units. Spatial contiguity would still be enforced through the corresponding adjacency graph, while the number of stored and evaluated dissimilarities would be substantially reduced.

\noindent\textbf{Hyperparameter calibration.}
Since posterior performance depends materially on the relative scales of the within-cluster and between-cluster dissimilarity models, hyperparameter calibration is an important part of the inferential pipeline. We therefore use a data-adaptive calibration scheme based on a pilot partition of the data, similar to the ones used in \citet{duan2021bayesian, natarajan2024cohesion}. Specifically, we first obtain a rough partition using a simple preliminary clustering procedure and use it to form provisional within-cluster and between-cluster dissimilarity sets. Gamma fits to these two sets are then used to calibrate the shape parameters $\delta_w$ and $\delta_b$ and to set the prior means of the latent rate parameters $\lambda_h$ and $\theta_{h\ell}$ through $(a_\lambda,b_\lambda)$ and $(a_\theta,b_\theta)$, so that the induced prior scales are aligned with the empirical within- and between-cluster dissimilarity distributions. The frailty precision $\kappa$ is calibrated through its direct interpretation as a control on node-level heterogeneity, since $\Var(w_i)=1/\kappa$ and $\mathrm{CV}(w_i)=\kappa^{-1/2}$. 
This yields a principled and reproducible calibration of the hyperparameters while preserving the Bayesian interpretation of the model. Full details of the calibration procedure, including the explicit mapping from pilot summaries to hyperparameter values, are provided in the Supplementary materials.

\section{Simulation Studies}
\label{sec:simulation}

\begin{figure}[t]
\centering
\includegraphics[width=\textwidth]{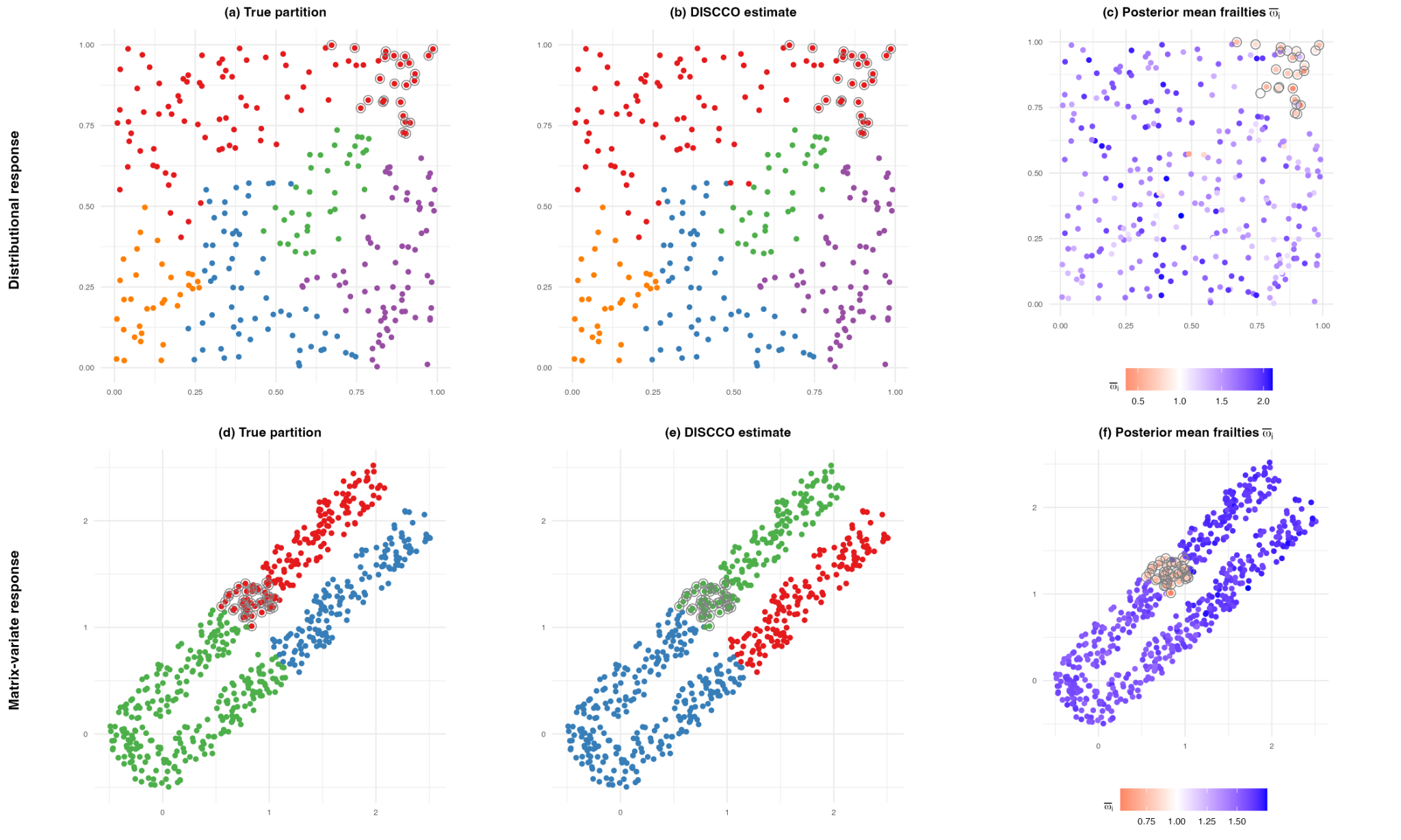}
\caption{Simulation results. \emph{Top row} (DGP1, distributional): (a)~true partition ($K=5$, patch circled), (b)~\textsc{DISCCO} VI estimate ($\hat{K}=5$), (c)~posterior mean frailties $\bar{w}_i$. \emph{Bottom row} (DGP2, matrix-variate on U-shape): (d)~true partition ($K=3$, patch circled), (e)~\textsc{DISCCO} VI estimate ($\hat{K}=3$), (f)~posterior mean frailties $\bar{w}_i$. In both scenarios, patch nodes (circled) receive lower frailties, confirming that the model identifies them as peripheral without creating spurious clusters.}
\label{fig:sim}
\end{figure}

\noindent\textbf{Data generation.} In DGP1 (distribution-valued response), each spatial unit is associated with an empirical income distribution. We generate \(n=300\) locations uniformly on \([0,1]^2\), construct a Delaunay adjacency graph, and form \(K=5\) contiguous true clusters (Figure~\ref{fig:sim}a). Following the contaminated Gamma construction of \cite{hu2023bayesian}, for unit \(i\) in cluster \(h\), the empirical distribution is based on an income sample of size \(n_i\) from the additive contamination model \(X_{ij}=Y_{ij}+B_{ij}Z_{ij}\), where \(Y_{ij}\sim \mathrm{Ga}_{\mathrm{sc}}(\alpha_h,50000)\), \(B_{ij}\sim \mathrm{Bernoulli}(0.05)\), \(Z_{ij}\sim \mathrm{Ga}_{\mathrm{sc}}(0.5,50000)\), \(\mathrm{Ga}_{\mathrm{sc}}(a,s)\) denotes a Gamma distribution with shape \(a\) and scale \(s\), and \((\alpha_1,\ldots,\alpha_5)=(0.80,1.10,1.50,2.00,2.60)\). To create localized sampling heterogeneity without changing the underlying cluster labels, we select a contiguous patch comprising \(25\%\) of the units in cluster~1 and set \(n_i=20\) for those units, while all remaining units have \(n_i=500\). 
Pairwise dissimilarities are computed using a discretized Wasserstein-2 distance, proportional to the Euclidean distance between empirical quantile vectors evaluated on $m=128$ equally spaced probability levels. This design tests whether a method can recover the true contiguous regions without fragmenting a spatially localized patch whose empirical distributions are noisier because of reduced sample size.

In DGP2 (matrix-variate response), $n=600$ locations are sampled uniformly within a non-convex U-shaped domain, yielding $K=3$ contiguous clusters: two on the arms and one at the bend (Figure~\ref{fig:sim}d). At each location $i$ in cluster $h$, the response is a $14 \times 14$ matrix drawn from a matrix normal distribution, $M_i \sim MN(\mu_h, U/\xi_i^2, V)$, where $\mu_h=\mathbf{a}_h\mathbf{1}^\top+\mathbf{1}\mathbf{b}_h^\top$ is a cluster-specific mean matrix. Specifically, for rows $r=1,\ldots,14$ and columns $c=1,\ldots,14$, $\mu_1(r,c)=25-\log r+t_c$, $\mu_2(r,c)=5-\log r+t_c$, and $\mu_3(r,c)=18-\frac{1}{2}\log r-s_c$, where $t_c$ is equally spaced from $0$ to $3$ and $s_c$ is equally spaced from $0$ to $2$. The baseline row covariance is $U=I_{14}+\mathbf{1}\mathbf{1}^\top/100$, and the column covariance is $V=I_{14}$. A contiguous patch of 25\% of cluster~1 receives inflated row covariance with $\xi_i=0.30$, corresponding to approximately $11.1\times$ covariance inflation, or $3.3\times$ standard-deviation-scale noise inflation, while all other locations have $\xi_i=1$. The pairwise dissimilarity is the Frobenius norm $\|M_i-M_j\|_F$. 

\noindent\textbf{Competitors.}
We compare \textsc{DISCCO} against cohesion--repulsion distance clustering \citep[RedClust;][]{natarajan2024cohesion} and Bayesian distance clustering \citep[BDC;][]{duan2021bayesian}, both of which are non-spatial distance-based clustering methods. RedClust is re-implemented in R with the same Gibbs sampler, Jain--Neal split-merge moves, and Pitman-type partition prior as the original Julia implementation; BDC is implemented by disabling the repulsion component, following the formulation in \citep{natarajan2024cohesion}. 
For DGP1 we additionally include Mclust \citep{scrucca2016mclust} and spectral clustering \citep{ng2001spectral}, both given the true \(K\). For DGP2 we additionally include a Dirichlet-process Gaussian mixture baseline for multivariate matrix summaries, implemented using \texttt{BNPmix} with discount parameter zero \citep{corradin2021bnpmix}; this baseline is applied to row-mean summaries of the matrix response augmented with standardized spatial coordinates and estimates \(K\) from the data. \textsc{DISCCO}, RedClust, BDC, and DP-MV estimate \(K\) throughout. Hyperparameters for \textsc{DISCCO} are set via the data-adaptive calibration of Section~\ref{sec:computation}.

\noindent\textbf{Results.}
We repeat the experiments 30 times for each DGP. 
Figure~\ref{fig:sim} displays the estimation results for both scenarios for a randomly selected experiment. In DGP1, \textsc{DISCCO} recovers $\hat{K}=5$ (panel~b), correctly retaining the patch within cluster~1. The frailty map (panel~c) confirms the mechanism: patch nodes receive lower posterior frailties ($\bar{w}_{\mathrm{patch}} = 0.745$, sd $= 0.250$) than non-patch nodes ($\bar{w}_{\mathrm{non\text{-}patch}} = 1.605$, sd $= 0.265$), absorbing the distance inflation from smaller sample sizes. In DGP2, \textsc{DISCCO} recovers $\hat{K}=3$ on the non-convex domain (panel~e), with patch nodes again receiving systematically lower frailties ($\bar{w}_{\mathrm{patch}} = 0.767 $, sd $= 0.067$; $\bar{w}_{\mathrm{non\text{-}patch}} = 1.604$, sd $= 0.065$; panel~f).

Table~\ref{tab:sim} summarizes results over 30 independent replications per scenario, evaluated using the adjusted Rand index \citep[ARI;][]{hubert1985comparing} and variation of information distance \citep[VI;][]{meilua2007comparing}. We also report mean wall-clock runtime, in seconds, over 5 benchmark runs of 2000 MCMC iterations, excluding hyperparameter calibration. In DGP1, \textsc{DISCCO} achieves the best partition recovery, with the lowest VI distance (0.116) and highest ARI (0.965). 
RedClust and BDC perform substantially worse, reflecting their tendency to fragment the noisy patch or otherwise misrepresent the spatial structure. In DGP2, 
\textsc{DISCCO} again attains the best numerical partition scores (VI \(=0.034\), ARI \(=0.991\)). 
In contrast, RedClust obtains a superficially high ARI (0.973) but massively overclusters, with \(\hat{K}_{\mathrm{mean}}=18.53\), a discrepancy more clearly reflected by its higher VI distance (0.154). BDC is more conservative, with \(\hat{K}_{\mathrm{mean}}=3.97\), but remains less accurate than DISCCO. The DP-MV baseline performs poorly in DGP2, with ARI \(=0.064\), VI \(=1.107\), and \(\hat{K}_{\mathrm{mean}}=2.17\), indicating that a non-contiguous mixture clustering model is not sufficient to recover the U-shaped spatial partition. The runtime comparisons show that, in our implementation, DISCCO is substantially faster than RedClust and BDC: for example, DISCCO is about \(16.6\times\) faster than RedClust in DGP1 and about \(27.6\times\) faster in DGP2. 

\begin{table}[t]
\centering
\caption{Simulation results: mean (sd) over 30 replications. Runtime is reported as the mean wall-clock time in seconds over 5 benchmark runs of 2000 MCMC iterations from each scenario. Methods marked with ${}^*$ are given the true $K$ as input parameters; all others estimate $K$ from the data.}
\label{tab:sim}
\small
\begin{tabular}{llcccc}
\toprule
& Method & VI distance & ARI & $\hat{K}$ & Runtime (s) \\
\midrule
\textbf{DGP1} & \textbf{DISCCO} & $\mathbf{0.116}$ (0.052) & $\mathbf{0.965}$ (0.016) & 5.90 (0.76) & 16.02 \\
& RedClust & 0.402 (0.067) & 0.714 (0.077) & 5.17 (0.59) & 266.02 \\
& BDC & 0.747 (0.134) & 0.683 (0.090) & 5.93 (0.87) & 206.71 \\
& Mclust${}^*$ & 0.511 (0.202) & 0.708 (0.116) & --- & --- \\
& Spectral${}^*$ & 0.506 (0.146) & 0.744 (0.089) & --- & --- \\
\midrule
\textbf{DGP2} & \textbf{DISCCO} & $\textbf{0.034}$ (0.027) & $\textbf{0.991}$ (0.008) & 3.17 (0.38) & 45.48 \\
& RedClust & 0.154 (0.059) & 0.973 (0.009) & 18.53 (6.10) & 1255.79 \\
& BDC & 0.146 (0.032) & 0.943 (0.015) & 3.97 (0.18) & 563.89 \\
& DP-MV & 1.107 (0.018) & 0.064 (0.006) & 2.17 (0.38) &  3.07\\
\bottomrule
\end{tabular}
\end{table}

\begin{figure}[t]
\centering
\includegraphics[width=0.85\linewidth]{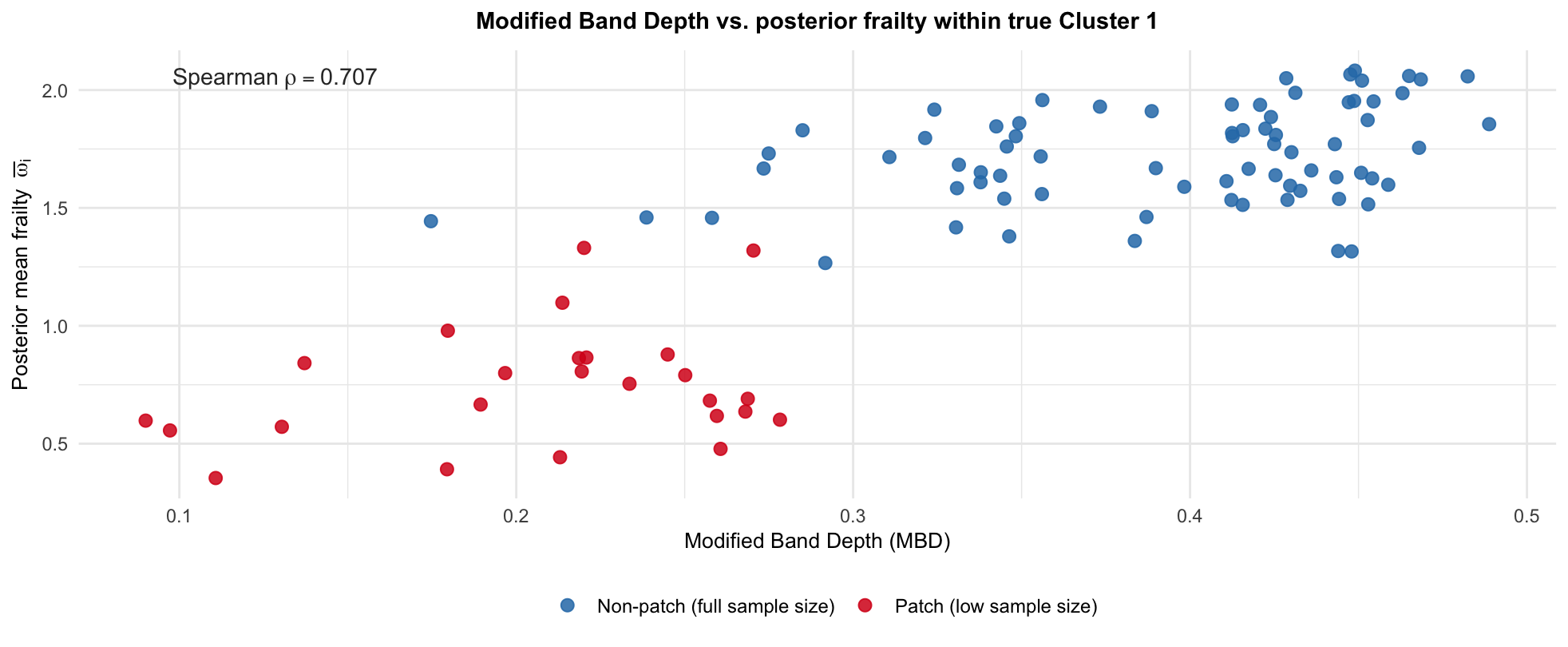}
\caption{External validation of frailty as centrality (DGP1, Cluster~1). $x$-axis: Modified Band Depth of each node's quantile function \citep{lopez2009concept}; $y$-axis: posterior mean frailty $\bar{w}_i$. Red = patch ($n_i=20$), blue = non-patch ($n_i=500$). Spearman $\rho = 0.707$.}
\label{fig:depth}
\end{figure}

\noindent\textbf{Frailty as a centrality measure: external validation.}
A key claim of \textsc{DISCCO} is that $w_i$ provides a model-based measure of within-cluster centrality. We validate this using Modified Band Depth \citep[MBD;][]{lopez2009concept}, a model-free functional depth that ranks curves by how frequently they lie within bands formed by other curves. Unlike DISCCO, MBD is a model-free depth measure and does not explicitly adjust for heterogeneous sampling variability. For a cluster-specific comparison, we compute the MBD of each node's empirical quantile function using the curves in the true Cluster~1 of DGP1 and plot it against the posterior mean frailty $\bar{w}_i$ (Figure~\ref{fig:depth}). The two quantities exhibit a strong positive monotone relationship (Spearman $\rho = 0.707$): patch nodes (red) concentrate in the lower-left (low depth, low $\bar{w}_i$) and non-patch nodes (blue) spread across the upper-right. This concordance is notable because MBD uses only the raw quantile curves with no model, while $\bar{w}_i$ is a posterior summary from the distance-based likelihood that never observes the quantile functions directly. The agreement confirms that the frailty recovers genuine within-cluster geometric centrality from the distance matrix alone.


\section{Real Data Analysis}
\label{sec:realdata}
We illustrate \textsc{DISCCO} on two spatial datasets in which each areal unit carries a complex object rather than a scalar response: a distribution-valued racial composition profile for Houston Census Block Groups and a matrix-valued cancer mortality profile for Western US counties. These analyses demonstrate both aspects of the proposed framework: recovery of spatially contiguous clusters from scientifically meaningful dissimilarities and posterior frailty summaries that identify central and peripheral units within each inferred region.

\subsection{Clustering of Racial Composition Distributions in Houston}
\label{sec:race}



\noindent\textbf{Data and setup.}
We apply \textsc{DISCCO} to racial composition data from $n=337$ CBGs in the urban core of Houston, Texas, including the Heights, Montrose, River Oaks, Museum District,
Midtown, Third Ward, East End, and Near Northside areas. At each CBG, the
observed object is a probability vector over four racial/ethnic categories:
non-Hispanic White (WH), African American (AA), Asian (AS), and Hispanic (HI),
obtained by normalizing American Community Survey counts to sum to one. Pairwise
distances are computed using the Hellinger distance, 
$H(p_i,p_j)= 2^{-\frac{1}{2}}\|\sqrt{p_i}-\sqrt{p_j}\|_2$,
and the spatial graph is constructed by connecting polygons sharing boundaries, with manual connectivity corrections applied when needed. We run \textsc{DISCCO} for
$20{,}000$ iterations with $10{,}000$ burn-in iterations, thinning by $5$. The reported partition is the point estimate
obtained using \texttt{salso} \citep{dahl2022search} based on VI minimization, and posterior mean
frailties $\bar w_i$ are used as within-cluster centrality summaries.

\begin{figure}[t]
  \centering
  \includegraphics[width=\linewidth]{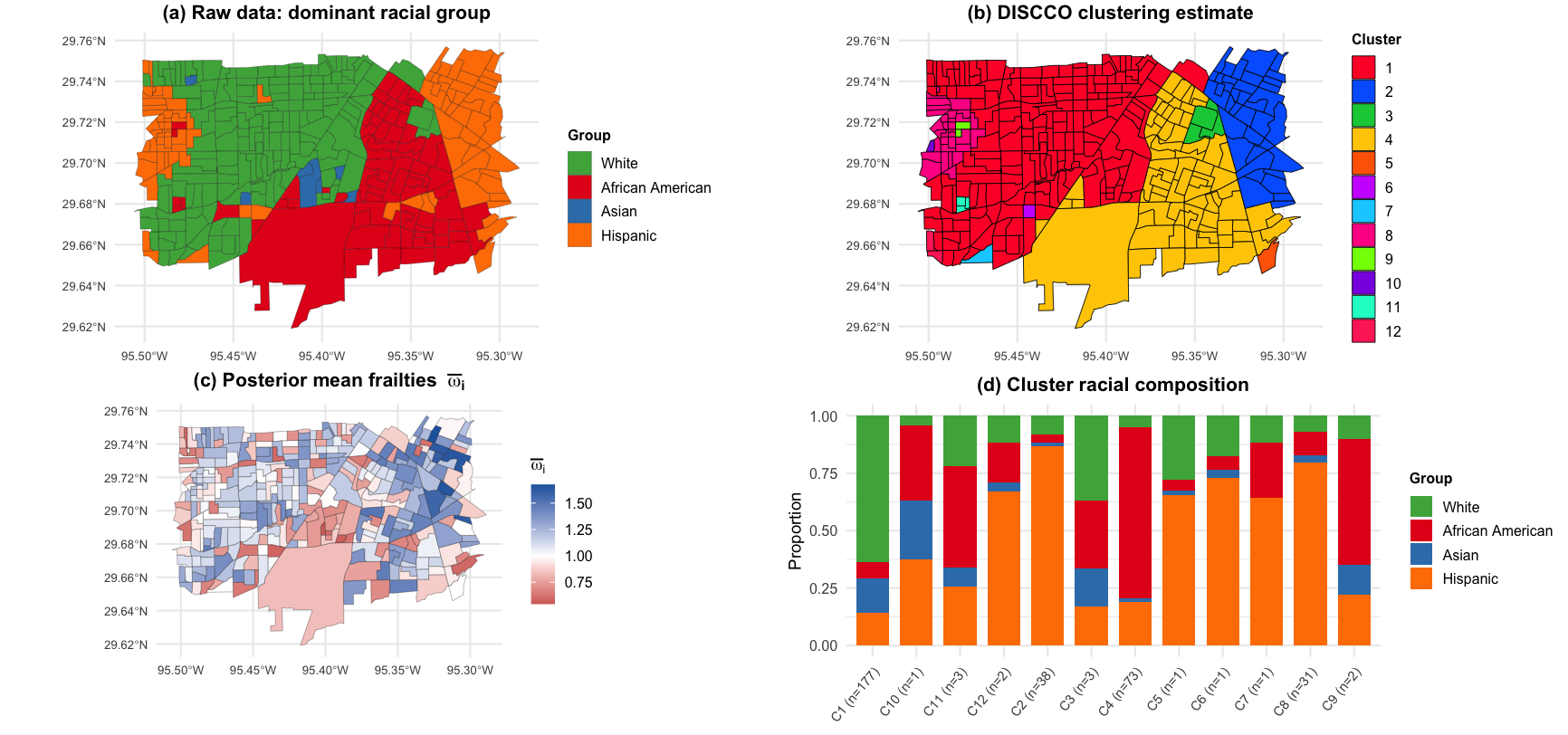}
  \caption{Houston racial composition clustering.
    \textbf{(a)} Dominant racial/ethnic group per CBG.
    \textbf{(b)} \textsc{DISCCO} VI-optimal partition with
    $\hat K=12$ contiguous regions.
    \textbf{(c)} Posterior mean frailties $\bar w_i$; larger values indicate
    CBGs that are more central within their inferred region's distributional
    geometry, while smaller values indicate more peripheral or transitional
    CBGs.
    \textbf{(d)} Mean racial composition of each inferred cluster.}
  \label{fig:houston_race}
\end{figure}

\noindent\textbf{Results.}
\textsc{DISCCO} recovers $\hat K=12$ spatially contiguous demographic regions
(Figure~\ref{fig:houston_race}b--d). The dominant-group map in
Figure~\ref{fig:houston_race}a is visually fragmented, since it records only the
largest racial/ethnic group in each CBG. In contrast, the \textsc{DISCCO}
partition uses the full four-category composition vector and aggregates nearby
CBGs into regions with coherent distributional profiles. The largest cluster, C1, spans much of the northwestern and central part of the
study region, broadly covering the Heights--Montrose--River Oaks--Museum
District--Midtown corridor. Its mean composition is White-dominant but mixed,
reflecting substantial Hispanic, Asian, and African American representation
within the same region. A second large region, C4, covers much of the southern
and southeastern areas, including the East
End, Second Ward, and Magnolia Park, and has a predominantly African American profile.
The northeastern cluster C2 is Hispanic-dominant, while the western cluster
C8 is also Hispanic-dominant. Several small
clusters and singletons capture localized composition profiles that are distinct
from their surrounding regions.

The posterior frailty map in Figure~\ref{fig:houston_race}c provides information
beyond the partition itself. Frailty centrality is interpreted only for non-singleton clusters; singleton frailties are prior-driven. 
Many low-frailty CBGs occur near boundaries between large regions or in locally heterogeneous areas, indicating transitional
composition profiles that need not form separate clusters. 

\begin{figure}[t]
\centering
\includegraphics[width=\linewidth]{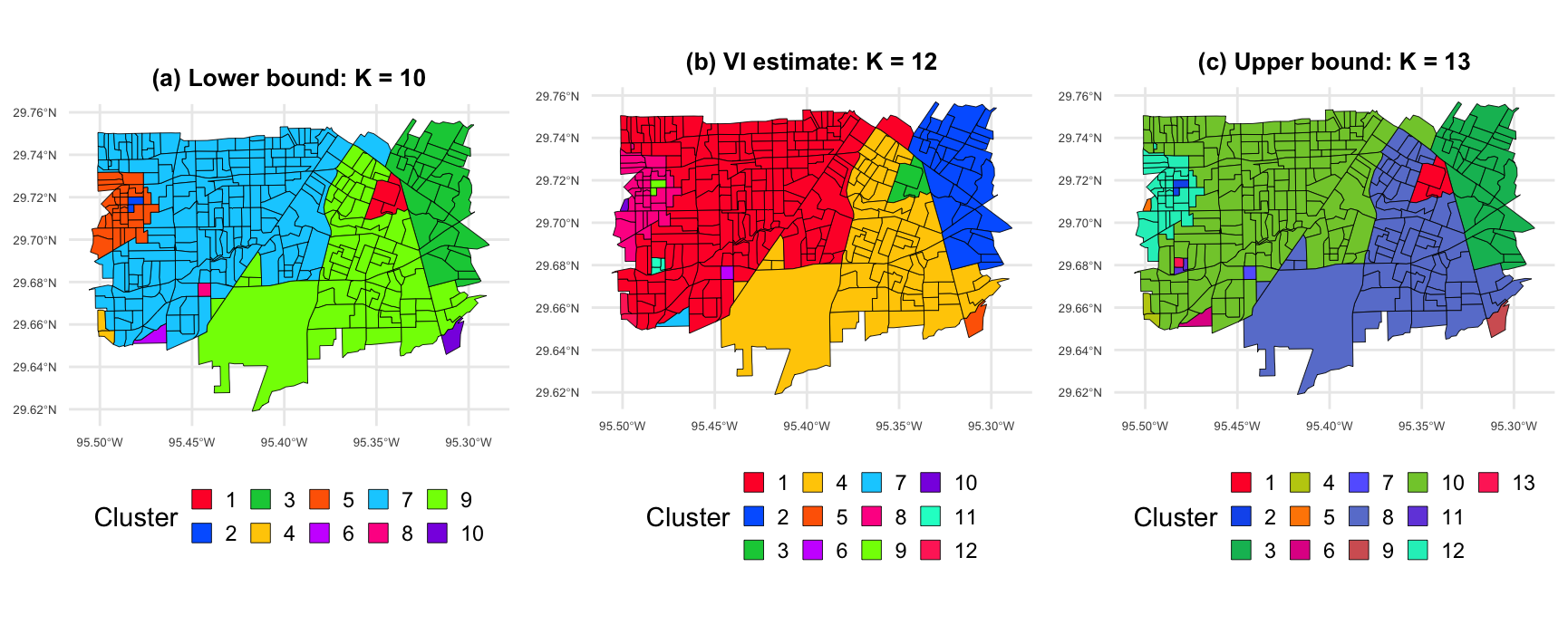}
\caption{Posterior credible-ball summaries for the Houston racial-composition analysis. The middle panel shows the VI-minimizing point estimate with \(\hat{K}=12\); the left and right panels show representative boundary partitions from the posterior credible ball. The large spatial regions are stable across the summaries, while uncertainty is concentrated in small localized regions and along boundaries between major clusters.}
\label{fig:houston_credible_ball}
\end{figure}

\begin{figure}[t]
\centering
\includegraphics[width=0.72\linewidth]{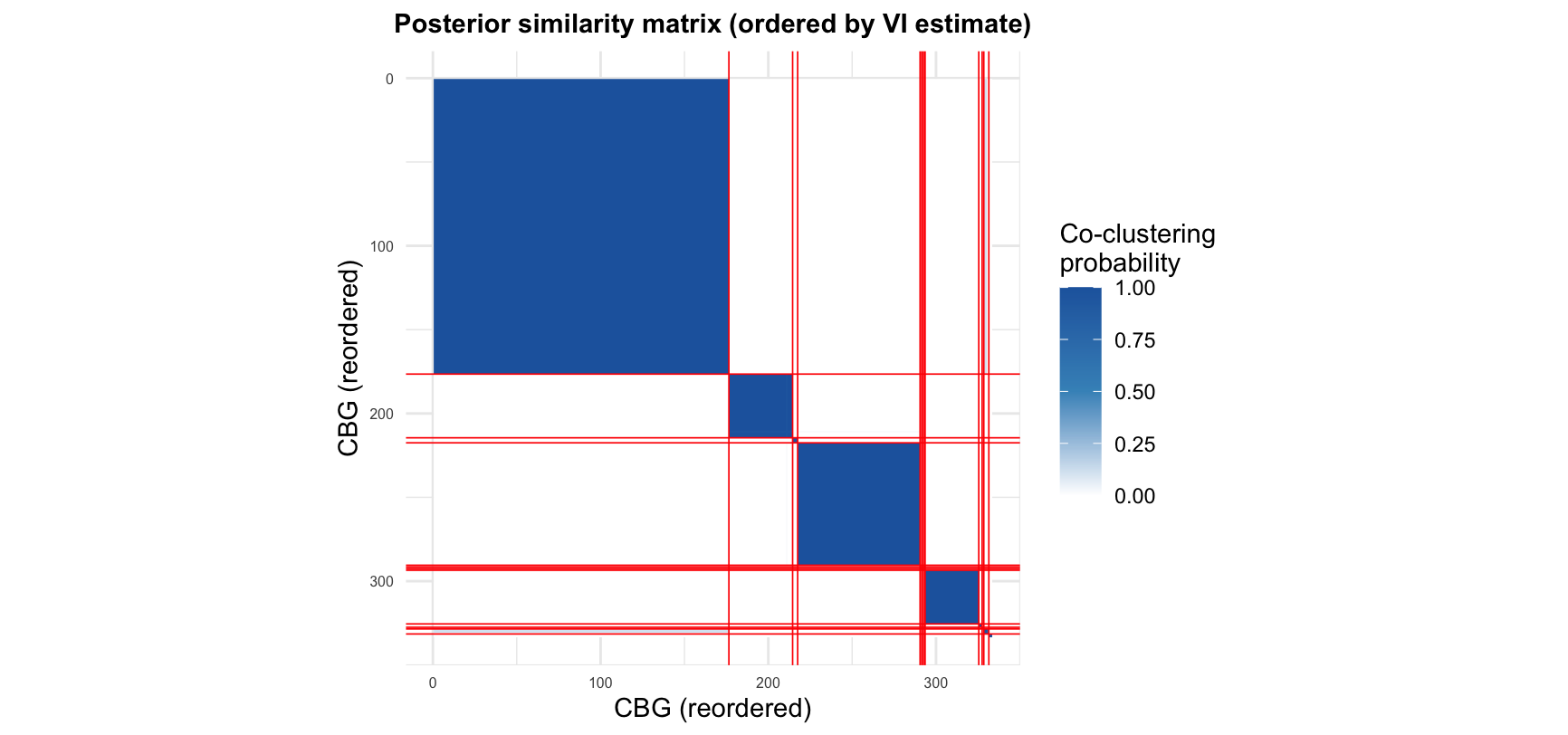}
\caption{Posterior similarity matrix for the Houston analysis, ordered by the VI point estimate. Entry \((i,j)\) is the posterior co-clustering probability \(P_{ij}=\Pr(i\sim j\mid d,G)\), and red lines mark the cluster boundaries in the VI estimate. The strong block-diagonal structure indicates high posterior support for the main inferred regions, with relatively little posterior mass assigned to cross-region co-clustering.}
\label{fig:houston_psm}
\end{figure}

Following the Bayesian clustering uncertainty summaries of
\citet{wade2018bayesian}, we assess posterior uncertainty for the Houston
racial-composition analysis using a $95\%$ credible ball around the
VI-minimizing point estimate $\hat{\rho}_{\mathrm{VI}}$ and the posterior
similarity matrix. The credible ball contains partitions within a specified VI
distance of $\hat{\rho}_{\mathrm{VI}}$ that together account for $95\%$ posterior
probability. We summarize its boundary using a finest representative partition,
which shows plausible splits of regions in $\hat{\rho}_{\mathrm{VI}}$, and a
coarsest representative partition, which shows plausible mergers of neighboring
regions. These extremal partitions characterize how far the clustering can
depart from the VI estimate while remaining within the credible ball. We also
consider the posterior co-clustering probabilities
$P_{ij}=\Pr(i\sim j\mid d,G)$. Figure~\ref{fig:houston_credible_ball} compares
$\hat{\rho}_{\mathrm{VI}}$ with the two credible-ball boundary partitions, while
Figure~\ref{fig:houston_psm} displays $P_{ij}$ after ordering the CBGs by
$\hat{\rho}_{\mathrm{VI}}$.

Figures~\ref{fig:houston_credible_ball}--\ref{fig:houston_psm} show that posterior uncertainty is localized rather than global. The VI point estimate and credible-ball boundary partitions recover the same broad spatial organization, with differences mainly involving small clusters and boundary CBGs. The posterior similarity matrix reinforces this conclusion: most within-cluster blocks have high co-clustering probability, while off-block probabilities are close to zero. 

The same posterior uncertainty assessment can be applied to the county-level
cancer-mortality analysis in Section~\ref{sec:cancer} without any modification. In that
setting, the credible-ball boundary partitions would identify plausible splits and mergers of the inferred mortality regions, while the posterior similarity matrix would show whether uncertainty is concentrated along county boundaries or reflects broader ambiguity in the regional assignments. Thus, these summaries provide a general post-processing tool for quantifying partition uncertainty from the DISCCO posterior, rather than being specific to the Houston application.

\subsection{Clustering Cancer Mortality Matrices Across Western US Counties}
\label{sec:cancer}

\noindent\textbf{Data and setup.}
We apply \textsc{DISCCO} to county-level cancer mortality profiles covering \(n=119\) counties across California, Nevada, Utah, and Arizona obtained from
the Global Health Data Exchange \citep{mokdad2017trends}. 
At each county, the observed object is an $8\times 8$ matrix whose rows correspond to eight cancer types (breast, lung, prostate, pancreatic, colorectal, liver, stomach, and kidney) and whose columns correspond to eight mortality-rate periods: 1976--1980, 1981--1985, 1986--1990, 1991--1995, 1996--2000, 2001--2005, 2006--2010, and 2011--2014. Thus each county is represented by a matrix-valued temporal cancer burden profile rather than by a scalar outcome. Pairwise dissimilarities are computed as Frobenius distances,
\[
d_{ij}=\|M_i-M_j\|_F,
\]
where \(M_i\) and \(M_j\) denote the cancer mortality matrices for counties \(i\) and \(j\). The spatial adjacency graph is constructed from county polygon contiguity, with small connectivity corrections applied when needed. We run \textsc{DISCCO} for \(30{,}000\) iterations with \(20{,}000\) burn-in iterations, thinning by 5. The reported partition is the VI-minimizing representative obtained using \texttt{salso}, and posterior mean frailties \(\bar{w}_i\) are used as within-cluster centrality summaries.

\begin{figure}[t]
  \centering
  \includegraphics[width=\linewidth]{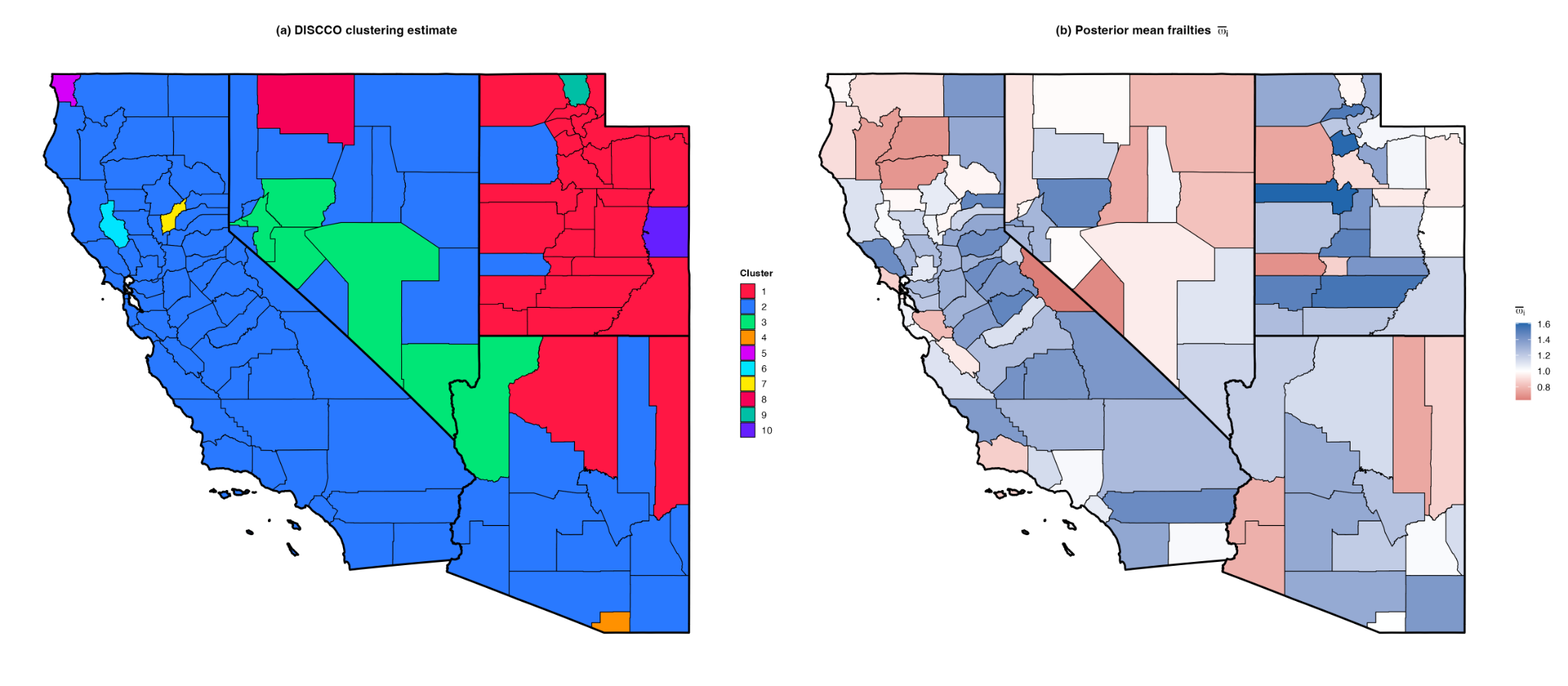}
  \caption{Cancer mortality matrix clustering across California, Nevada, Utah, and Arizona. \textbf{(a)} \textsc{DISCCO} VI-optimal partition with \(\hat{K}=10\) contiguous regions; thick black lines indicate state boundaries. \textbf{(b)} Posterior mean frailties \(\bar{w}_i\); larger values indicate counties that are more central within their inferred region's cancer-mortality matrix geometry, while smaller values indicate more peripheral or transitional counties.}
  \label{fig:cancer}
\end{figure}

\noindent\textbf{Results.}
\textsc{DISCCO} recovers \(\hat{K}=10\) spatially contiguous regions (Figure~\ref{fig:cancer}). The largest region, C2, covers much of California and Arizona, suggesting that these counties share broadly similar cancer-by-time mortality profiles. A second large region, C1, covers Utah together with adjacent counties in eastern Nevada and northern Arizona. This region is separated from the broader western baseline by its full \(8 \times 8\) mortality matrix, indicating a coherent temporal cancer-profile pattern across multiple cancer types. Several additional regions capture localized structure within Nevada and along state-border areas, while a small number of localized clusters identify counties or groups of counties whose mortality matrices are distinct from their geographic surroundings.

The partition should be interpreted as a clustering of full multivariate temporal profiles rather than as a map of any single cancer type or single year. For example, the Utah-centered region C1 is not defined by one isolated mortality rate, but by a joint pattern across eight cancers and eight time points. Similarly, the northern and interior Nevada regions separate from neighboring counties because their overall matrix-valued profiles are more similar internally than to adjacent large regions. The partition preserves the broad western, Utah-centered, and Nevada/interior structures while absorbing some highly localized variation into neighboring regions. These patterns may reflect differences in demographic composition, population size, rurality, exposure history, or healthcare access, but the clustering model itself uses only the mortality matrices and the county adjacency graph. We therefore view the recovered regions as descriptive summaries of spatially coherent cancer-mortality profiles rather than as causal explanations of cancer risk.

The posterior frailty map in Figure~\ref{fig:cancer}b provides information beyond the discrete partition. Counties with large \(\bar{w}_i\) are central within their inferred region, meaning that their full cancer-by-time mortality matrices have relatively small within-cluster distances after accounting for the region-level compactness parameter. Counties with small \(\bar{w}_i\) are more peripheral, indicating that their matrices are less representative of the fitted regional profile. Several low-frailty counties occur near boundaries between large regions or in geographically isolated areas, suggesting transitional or atypical mortality profiles that need not form separate clusters. 

\section{Conclusion}
\label{sec:conclusion}
We introduced \textsc{DISCCO}, a Bayesian distance-based framework for clustering spatially indexed complex objects into contiguous regions. The model specifies a new hierarchical distance likelihood whose node frailties account for shared-node dependence and provide centrality/peripherality summaries. Latent clusters are modeled by a spanning-tree-based random graph partition prior for spatial contiguity, which leads to an efficient partition update algorithm and a Gibbs tree update sampler. Simulations show that \textsc{DISCCO} accurately recovers contiguous regions and avoids fragmenting noisy spatial patches, while the real-data analyses illustrate its broad applicability in practice for various types of object data. Moving forward, there are several promising avenues to address the limitations of DISCCO. The current implementation of DISCCO only handles moderate to large-sized spatial graphs. There are several computational strategies to scale DISCCO to massive graphs, including sparse-distance representations, neighborhood-restricted likelihoods, and graph dimension reduction by node blocking.  The current method addresses spatial clustering only and does not incorporate covariates, and extensions to spatio-temporal clustering and covariate-dependent clustering are promising directions to enable broader applications. Moreover, the DISCCO framework can be naturally extended beyond spatial settings to accommodate more general object data observed on networks, such as user activity profiles on social networks or load curves on the power grid. Establishing posterior contraction theory for contiguous partition recovery is another important direction for future work~\citep{huang2025consistent}.













\bibliography{bibliography.bib}

\newpage
\appendix

\section{Proof of Proposition 1}
\label{app:proof_prop1}

We prove the covariance statements in Proposition~1. Throughout, all Gamma
distributions are in shape--rate parameterization. We use the following inverse
moments. If $X\sim\Ga(a,b)$ with $a>2$, then
\[
\E(X^{-1})=\frac{b}{a-1},
\qquad
\E(X^{-2})=\frac{b^2}{(a-1)(a-2)},
\]
and hence
\[
\Var(X^{-1})
=
\E(X^{-2})-\{\E(X^{-1})\}^2
=
\frac{b^2}{(a-1)^2(a-2)}.
\]
For the node frailties $w_i\sim\Ga(\kappa,\kappa)$ with $\kappa>2$, define
\[
\nu_1=\E(w_i^{-1})=\frac{\kappa}{\kappa-1},
\qquad
\nu_2=\E(w_i^{-2})
=
\frac{\kappa^2}{(\kappa-1)(\kappa-2)}.
\]
Then
\[
\nu_2-\nu_1^2
=
\Var(w_i^{-1})
=
\frac{\kappa^2}{(\kappa-1)^2(\kappa-2)}
>0.
\]
Similarly, for $\lambda_h\sim\Ga(a_\lambda,b_\lambda)$, define
\[
\mu_{\lambda,1}=\E(\lambda_h^{-1})
=
\frac{b_\lambda}{a_\lambda-1},
\qquad
\mu_{\lambda,2}=\E(\lambda_h^{-2})
=
\frac{b_\lambda^2}{(a_\lambda-1)(a_\lambda-2)}.
\]
Since $a_\lambda>2$,
\[
\mu_{\lambda,2}-\mu_{\lambda,1}^2
=
\Var(\lambda_h^{-1})
=
\frac{b_\lambda^2}{(a_\lambda-1)^2(a_\lambda-2)}
>0.
\]

\noindent\textbf{Within-cluster covariances.}
Fix a block $S_h$ and suppose $i,j,k,\ell\in S_h$ are distinct. Conditional on
$(\lambda_h,w)$, the within-cluster distances are independent and
\[
\E(d_{ij}\mid \lambda_h,w)
=
\frac{\delta_w}{\lambda_h w_iw_j}.
\]
Therefore
\[
\E(d_{ij})
=
\delta_w \mu_{\lambda,1}\nu_1^2.
\]

First, consider two disjoint within-cluster distances, $d_{ij}$ and $d_{k\ell}$.
Because the two distances share only the block-level compactness parameter
$\lambda_h$,
\[
\E(d_{ij}d_{k\ell})
=
\delta_w^2
\E(\lambda_h^{-2})
\E(w_i^{-1})\E(w_j^{-1})\E(w_k^{-1})\E(w_\ell^{-1})
=
\delta_w^2\mu_{\lambda,2}\nu_1^4.
\]
Thus
\[
\Cov(d_{ij},d_{k\ell})
=
\delta_w^2\nu_1^4
\left(\mu_{\lambda,2}-\mu_{\lambda,1}^2\right)
=
\delta_w^2\nu_1^4\Var(\lambda_h^{-1})
>0.
\]

Now consider two within-cluster distances sharing one endpoint, $d_{ij}$ and
$d_{ik}$. Conditional independence again gives
\[
\E(d_{ij}d_{ik})
=
\delta_w^2
\E(\lambda_h^{-2})
\E(w_i^{-2})\E(w_j^{-1})\E(w_k^{-1})
=
\delta_w^2\mu_{\lambda,2}\nu_2\nu_1^2.
\]
Therefore
\[
\Cov(d_{ij},d_{ik})
=
\delta_w^2
\left(
\mu_{\lambda,2}\nu_2\nu_1^2
-
\mu_{\lambda,1}^2\nu_1^4
\right).
\]
Subtracting the disjoint-pair covariance yields
\[
\Cov(d_{ij},d_{ik})-\Cov(d_{ij},d_{k\ell})
=
\delta_w^2\mu_{\lambda,2}\nu_1^2
\left(\nu_2-\nu_1^2\right)
=
\delta_w^2\mu_{\lambda,2}\nu_1^2\Var(w_i^{-1})
>0.
\]
Hence
\[
\Cov(d_{ij},d_{ik})
>
\Cov(d_{ij},d_{k\ell})
>
0.
\]
This proves the first statement.

\noindent\textbf{Between-cluster covariances for a fixed block pair.}
Now fix two distinct blocks $S_h$ and $S_\ell$. For $i,i'\in S_h$ and
$j,j'\in S_\ell$, the between-cluster distances associated with the unordered
block pair $\{S_h,S_\ell\}$ share the same latent rate $\theta_{h\ell}$, where
\[
\theta_{h\ell}\sim\Ga(a_\theta,b_\theta).
\]
Let
\[
\mu_{\theta,1}
=
\E(\theta_{h\ell}^{-1})
=
\frac{b_\theta}{a_\theta-1},
\qquad
\mu_{\theta,2}
=
\E(\theta_{h\ell}^{-2})
=
\frac{b_\theta^2}{(a_\theta-1)(a_\theta-2)}.
\]
For two distinct cross-block distances $d_{ij}$ and $d_{i'j'}$, conditional
independence gives
\[
\E(d_{ij})
=
\delta_b\mu_{\theta,1},
\qquad
\E(d_{ij}d_{i'j'})
=
\delta_b^2\mu_{\theta,2}.
\]
Thus
\[
\Cov(d_{ij},d_{i'j'})
=
\delta_b^2
\left(
\mu_{\theta,2}-\mu_{\theta,1}^2
\right)
=
\delta_b^2\Var(\theta_{h\ell}^{-1})
=
\delta_b^2
\frac{b_\theta^2}{(a_\theta-1)^2(a_\theta-2)}
>0.
\]
This expression depends only on the unordered block pair through the shared
parameter $\theta_{h\ell}$ and the common hyperparameters, not on the particular
nodes $i,i',j,j'$. If the two distances are identical, the covariance is the
variance of that distance, which is also positive.

\noindent\textbf{Between-cluster distances from different block pairs.}
Consider two between-cluster distances associated with different unordered
block pairs, for example one distance connecting $S_h$ and $S_\ell$ and another
connecting $S_r$ and $S_s$, with
\[
\{h,\ell\}\neq \{r,s\}.
\]
The first distance depends only on $\theta_{h\ell}$, while the second depends
only on $\theta_{rs}$. The parameters $\theta_{h\ell}$ and $\theta_{rs}$ are
independent a priori, and the distances are conditionally independent given the
latent rates. Therefore the two distances are marginally independent.

\noindent\textbf{Independence of within- and between-cluster distances.}
Finally, a within-cluster distance $d_{ij}$ with $i,j\in S_h$ depends on
$\lambda_h$ and the frailties $w_i,w_j$. A between-cluster distance
$d_{rs}$ with $r\in S_a$ and $s\in S_b$, $a\neq b$, depends only on the
between-cluster rate $\theta_{ab}$. The latent variables
\[
\lambda_h,\quad w_1,\ldots,w_n,\quad \theta_{ab}
\]
are mutually independent under the prior, and the distances are conditionally
independent given these latent variables. Hence any within-cluster distance is
marginally independent of any between-cluster distance.

Combining the four arguments proves Proposition~1.

\section{Hyperparameter Calibration}
\label{app:hyperparameters}

This section describes the default calibration used for the likelihood
hyperparameters
\[
(\delta_w,a_\lambda,b_\lambda,\delta_b,a_\theta,b_\theta,\kappa)
\]
and the partition-prior hyperparameters.   The calibration is data-adaptive and only requires the
empirical scale of the observed dissimilarities and a pilot partition. Posterior
inference is then performed using the full \textsc{DISCCO} model described in
the main paper. Our numerical results indicate that this calibration strategy works well in practice. Below, we provide detailed calibration steps.

\subsection{Pilot distance splits}

Let $
\rho_{\mathrm{pilot}}=\{C_1,\ldots,C_{K_{\mathrm{pilot}}}\}$
denote a preliminary partition of the observed units. This pilot
partition is used only to calibrate the scale of the Gamma likelihoods. In the reported experiments, it is obtained by applying $k$-means to the available
finite-dimensional representation of the objects, with $K_{\mathrm{pilot}}$ selected
by an elbow criterion. When only a dissimilarity matrix is available, a
dissimilarity-based pilot method such as hierarchical clustering or $k$-medoids
may be used instead. The quantity \(K_{\mathrm{pilot}}\) is used only in the data-adaptive calibration step. Specifically, pilot partitions may be used to form provisional within-cluster and between-cluster distance sets for calibrating the likelihood hyperparameters \((\delta_w,\delta_b,a_\lambda,b_\lambda,a_\theta,b_\theta)\). This pilot value does not define the support of the posterior distribution of \(K\), and it is not a hard upper bound on the number of clusters sampled by the MCMC algorithm. Thus, posterior summaries such as \(\hat K\) may exceed \(K_{\mathrm{pilot}}\) if supported by the distance matrix and spatial graph.

Using $\rho_{\mathrm{pilot}}$, form the empirical within-cluster and
between-cluster distance sets $\mathcal A
=
\{d_{ij}: i,j \text{ belong to the same pilot cluster}\}$, and $\mathcal B
=
\{d_{ij}: i,j \text{ belong to different pilot clusters}\}$.

Let
$
\bar d_{\mathcal A}
=
\frac{1}{|\mathcal A|}
\sum_{a\in\mathcal A} a$,
$\bar d_{\mathcal B}
=
\frac{1}{|\mathcal B|}
\sum_{b\in\mathcal B} b
$
denote the corresponding empirical means, respectively.

\subsection{Calibrating the Gamma distance shapes}

We fit Gamma distributions in shape--rate form to the two pilot distance sets: $\mathcal A \approx \Ga(\hat\delta_w,\hat r_w),
\mathcal B \approx \Ga(\hat\delta_b,\hat r_b)$.

Equivalently, one may use method-of-moments estimates $\hat\delta_w
=
\frac{\bar d_{\mathcal A}^2}{s_{\mathcal A}^2},
\hat r_w
=
\frac{\bar d_{\mathcal A}}{s_{\mathcal A}^2}
=
\frac{\hat\delta_w}{\bar d_{\mathcal A}}$, and $\hat\delta_b
=
\frac{\bar d_{\mathcal B}^2}{s_{\mathcal B}^2},
\hat r_b
=
\frac{\bar d_{\mathcal B}}{s_{\mathcal B}^2}
=
\frac{\hat\delta_b}{\bar d_{\mathcal B}}$, where $s_{\mathcal A}^2$ and $s_{\mathcal B}^2$ are the empirical variances of
the pilot within-cluster and between-cluster distance sets. We then set $\delta_w=\hat\delta_w$. For the between-cluster component, we use repulsive calibration
$\delta_b=\max\{\hat\delta_b,1+\epsilon\}$, for a small fixed $\epsilon>0$. This ensures that the between-cluster Gamma density vanishes at the origin.

\subsection{Calibrating the priors on compactness and separation rates}

The priors on the cluster compactness and block-pair separation rates are $\lambda_h\stackrel{iid}{\sim}\Ga(a_\lambda,b_\lambda),
\qquad
\theta_{h\ell}\stackrel{iid}{\sim}\Ga(a_\theta,b_\theta)$. We choose these priors so that their means match the pilot fitted rate scales,
while their effective prior strength is of order $O(n)$ rather than $O(n^2)$.
Let $n_{\mathrm{pseudo}}=n$. We set $a_\lambda=\delta_w n_{\mathrm{pseudo}},
\qquad
b_\lambda=n_{\mathrm{pseudo}}\bar d_{\mathcal A}$, and $a_\theta=\delta_b n_{\mathrm{pseudo}},
\qquad
b_\theta=n_{\mathrm{pseudo}}\bar d_{\mathcal B}$. Then $\E(\lambda_h)
=
\frac{a_\lambda}{b_\lambda}
=
\frac{\delta_w}{\bar d_{\mathcal A}}$, and $\E(\theta_{h\ell})
=
\frac{a_\theta}{b_\theta}
=
\frac{\delta_b}{\bar d_{\mathcal B}}$.

Thus the prior means are centered at the fitted within-cluster and
between-cluster distance-rate scales. The use of $n_{\mathrm{pseudo}}=n$ keeps
the prior informative enough to stabilize computation while remaining weak relative to the nominal \(O(n^2)\) pairwise distance contributions in the likelihood available in moderate and
large datasets.





\subsection{Calibrating frailty scale $\kappa$}

The within-cluster likelihood depends on the product $\lambda_h w_i w_j $.

Without a scale normalization on the frailties, the cluster-level compactness
parameter $\lambda_h$ and the node-level frailties would not be separately
identified. We therefore use $w_i \stackrel{iid}{\sim} \Ga(\kappa,\kappa)$, so that $\E(w_i)=1,
\Var(w_i)=\frac{1}{\kappa},
\mathrm{CV}(w_i)=\frac{1}{\sqrt{\kappa}}$. This centers the frailties at the neutral value one, leaving $\lambda_h$ to
control the cluster-level distance scale and $w_i$ to encode node-specific
departures from that scale. With the frailties centered at the neutral value \(w_i=1\), these choices center the prior rate scales so that the implied conditional within- and between-cluster distance scales agree with the pilot empirical means.

A convenient calibration is to specify a target prior coefficient of variation
$c_w=\mathrm{CV}(w_i)$ and set $\kappa=c_w^{-2}$. For example, $c_w=0.20 \quad \Rightarrow \quad \kappa=25,
c_w=\frac{1}{3} \quad \Rightarrow \quad \kappa=9,
c_w=0.50 \quad \Rightarrow \quad \kappa=4$. In practice, we recommend fitting the model over a small grid, such as $\kappa\in\{4,9,25\}$, and checking the sensitivity of posterior partitions and frailty summaries. In implementation, we restrict the calibrated values to satisfy \(a_\lambda>2\), \(a_\theta>2\), and \(\kappa>2\), matching the moment conditions used in Proposition~1.

\subsection{Partition-prior calibration}

The tree-cut prior used for posterior inference places a truncated geometric prior on the number of clusters over the full partition support, $p(K)\propto \eta^{K-1},
\qquad
K\in\{1,\ldots,n\}$. Equivalently, for \(\eta\neq 1\), $p(K=k)
=
\frac{(1-\eta)\eta^{k-1}}
     {1-\eta^{n}},
\qquad
k=1,\ldots,n$, and for \(\eta=1\), the prior is uniform on \(\{1,\ldots,n\}\). The parameter \(\eta\) controls prior preference for the number of clusters: values \(\eta<1\) penalize larger \(K\), while values \(\eta>1\) favor larger \(K\). In the reported experiments we take \(\eta<1\), giving a weak regularization toward fewer clusters while still allowing the posterior number of clusters to exceed any pilot value used during calibration. For reference, under the actual prior support \(\{1,\ldots,n\}\), the prior mean is $\E(K)
=
1+
\frac{
\eta\{1-n\eta^{n-1}
+(n-1)\eta^{n}\}
}{
(1-\eta)(1-\eta^{n})
},
\qquad \eta\neq 1$, and $\E(K)=\frac{n+1}{2},
\qquad \eta=1$. Sensitivity to \(\eta\) can be assessed by comparing the posterior distribution of \(K\), posterior co-clustering matrices, and representative partitions across several values of the geometric-prior decay parameter.

\subsection{Summary of the default calibration}

The default calibration proceeds as follows:
\begin{enumerate}
    \item Obtain a pilot partition $\rho_{\mathrm{pilot}}$ from the observed units.
    \item Construct the pilot within-cluster and between-cluster distance sets
    $\mathcal A$ and $\mathcal B$.
    \item Fit Gamma shapes $\hat\delta_w$ and $\hat\delta_b$ to
    $\mathcal A$ and $\mathcal B$.
    \item Set $\delta_w=\hat\delta_w$ and set
    $\delta_b=\max\{\hat\delta_b,1+\epsilon\}$.
    \item Set $a_\lambda=\delta_w n,
    b_\lambda=n\bar d_{\mathcal A},
    a_\theta=\delta_b n,
    b_\theta=n\bar d_{\mathcal B}$.
    \item Choose $\kappa=c_w^{-2}$ from a target frailty coefficient of
    variation $c_w$, and assess sensitivity over a small grid of $\kappa$ values.
    \item Choose \(\eta\) to give a weakly regularizing prior on the number of clusters, and assess sensitivity of posterior partitions and \(K\) to \(\eta\). The pilot value \(K_{\mathrm{pilot}}\) is used only for likelihood calibration and does not restrict the posterior support of \(K\).
\end{enumerate}
This calibration places the likelihood on the empirical scale of the observed
dissimilarities while preserving the intended separation between cluster-level
compactness, between-cluster separation, and node-level frailty heterogeneity.

In all reported experiments, we fixed the frailty coefficient of variation at
$\operatorname{CV}(w_i)=1/2$, corresponding to $\kappa=4$, and fixed the
partition-prior decay parameter at $\eta=0.8$. These values were used as default
weakly informative settings across all simulation and real-data analyses, rather
than selected separately for each dataset. The choice
$\operatorname{CV}(w_i)=1/2$ allows moderate node-level heterogeneity while
keeping the frailties centered at the neutral value 1, and $\eta=0.8$ provides
mild prior regularization against excessive fragmentation. 

\section{MCMC Details}
\label{app:mcmc}

This section gives the collapsed likelihood factors and the Markov chain Monte
Carlo updates used for posterior inference. We use the notation of the main
paper throughout. The Markov state is
\[
\{T,B,K,\rho, w \}
\]
where $T$ is a spanning tree of the spatial graph $G$, $B\subseteq E(T)$ is the
cut set, $K=|B|+1$, $\rho=\rho(T,B)=\{S_1,\ldots,S_K\}$ is the induced connected
partition, and $w=(w_1,\ldots,w_n)$ are the node frailties. The compactness
parameters $\lambda_h$ and separation parameters $\theta_{h\ell}$ are integrated
out analytically, while the frailties are retained in the Markov state.

\subsection{Collapsed within-cluster factor}

Fix a block $S\subseteq V$ and define
\[
n_S=|S|,
\qquad
m_S=\binom{n_S}{2},
\qquad
A_S=\sum_{i<j:\,i,j\in S}\log d_{ij},
\]
and
\[
D_S^{(w)}=\sum_{i<j:\,i,j\in S}w_iw_jd_{ij}.
\]
Conditional on $\lambda_S$ and the frailties, the within-cluster contribution is
\[
\prod_{i<j:\,i,j\in S}
f(d_{ij}\mid \lambda_S,w_i,w_j)
=
\Gamma(\delta_w)^{-m_S}
\exp\{(\delta_w-1)A_S\}
\lambda_S^{\delta_w m_S}
\left\{\prod_{i\in S}w_i^{\delta_w(n_S-1)}\right\}
\exp\{-\lambda_S D_S^{(w)}\}.
\]
Using the prior $\lambda_S\sim\Ga(a_\lambda,b_\lambda)$ in shape--rate
parameterization, the collapsed within-block factor is
\[
W_w(S)
=
\frac{b_\lambda^{a_\lambda}}{\Gamma(a_\lambda)}
\Gamma(\delta_w)^{-m_S}
\exp\{(\delta_w-1)A_S\}
\left\{\prod_{i\in S}w_i^{\delta_w(n_S-1)}\right\}
\frac{\Gamma(a_\lambda+\delta_w m_S)}
     {(b_\lambda+D_S^{(w)})^{a_\lambda+\delta_w m_S}}.
\]
Equivalently,
\begin{align*}
\log W_w(S)
={}&
a_\lambda\log b_\lambda-\log\Gamma(a_\lambda)
-m_S\log\Gamma(\delta_w)
+(\delta_w-1)A_S  \\
&+\delta_w(n_S-1)\sum_{i\in S}\log w_i
+\log\Gamma(a_\lambda+\delta_w m_S)
-(a_\lambda+\delta_w m_S)\log(b_\lambda+D_S^{(w)}).
\end{align*}
For a singleton block, $m_S=0$, $A_S=0$, and $D_S^{(w)}=0$, so $W_w(S)=1$.

\subsection{Collapsed between-cluster factor}

For two distinct blocks $S,T\subseteq V$, define
\[
r_{ST}=|S||T|,
\qquad
A_{ST}=\sum_{i\in S}\sum_{j\in T}\log d_{ij},
\qquad
D_{ST}=\sum_{i\in S}\sum_{j\in T}d_{ij}.
\]
Conditional on $\theta_{ST}$, the between-cluster contribution is
\[
\prod_{i\in S}\prod_{j\in T}g(d_{ij}\mid \theta_{ST})
=
\Gamma(\delta_b)^{-r_{ST}}
\exp\{(\delta_b-1)A_{ST}\}
\theta_{ST}^{\delta_b r_{ST}}
\exp\{-\theta_{ST}D_{ST}\}.
\]
Using the prior $\theta_{ST}\sim\Ga(a_\theta,b_\theta)$, the collapsed
between-block factor is
\[
R(S,T)
=
\frac{b_\theta^{a_\theta}}{\Gamma(a_\theta)}
\Gamma(\delta_b)^{-r_{ST}}
\exp\{(\delta_b-1)A_{ST}\}
\frac{\Gamma(a_\theta+\delta_b r_{ST})}
     {(b_\theta+D_{ST})^{a_\theta+\delta_b r_{ST}}}.
\]
Equivalently,
\begin{align*}
\log R(S,T)
={}&
a_\theta\log b_\theta-\log\Gamma(a_\theta)
-r_{ST}\log\Gamma(\delta_b)
+(\delta_b-1)A_{ST} \\
&+\log\Gamma(a_\theta+\delta_b r_{ST})
-(a_\theta+\delta_b r_{ST})\log(b_\theta+D_{ST}).
\end{align*}

\subsection{Partially collapsed posterior}

After integrating out $\lambda_1,\ldots,\lambda_K$ and
$\{\theta_{h\ell}:1\leq h<\ell\leq K\}$, the posterior target is
\[
\pi(T,B,K,\rho,w\mid d,G)
\propto
\pi_0(T,B,K,\rho\mid G)
p(w\mid \kappa)
\prod_{h=1}^K W_w(S_h)
\prod_{1\leq h<\ell\leq K}R(S_h,S_\ell),
\]
where
\[
p(w\mid \kappa)
=
\prod_{i=1}^n
\frac{\kappa^\kappa}{\Gamma(\kappa)}
w_i^{\kappa-1}\exp(-\kappa w_i).
\]
Define the collapsed distance score
\[
\ell(\rho,w)
=
\sum_{h=1}^K\log W_w(S_h)
+
\sum_{1\leq h<\ell\leq K}\log R(S_h,S_\ell).
\]
Then the full collapsed log target can be written as
\[
\Phi(T,B,K,\rho,w)
=
\log \pi_0(T,B,K,\rho\mid G)
+
\log p(w\mid \kappa)
+
\ell(\rho,w).
\]

\subsection{Conditional distributions of the collapsed rates}

Although the sampler does not require sampling $\lambda_h$ or $\theta_{h\ell}$,
their posterior full conditionals are available in closed form. Given
$(\rho,w)$,
\[
\lambda_h\mid \rho,w,d
\sim
\Ga\left(
a_\lambda+\delta_w m_{S_h},
\,
b_\lambda+D_{S_h}^{(w)}
\right),
\qquad h=1,\ldots,K.
\]
Similarly, for $1\leq h<\ell\leq K$,
\[
\theta_{h\ell}\mid \rho,d
\sim
\Ga\left(
a_\theta+\delta_b r_{S_hS_\ell},
\,
b_\theta+D_{S_hS_\ell}
\right).
\]
These draws may be used for posterior summaries but are not part of the
collapsed partition sampler.

\subsection{Local identities}

The following identities are useful for local score updates. Suppose a block
$S$ is split into two nonempty connected pieces $S_1$ and $S_2$. Then
\[
A_S=A_{S_1}+A_{S_2}+A_{S_1S_2},
\]
and
\[
D_S^{(w)}
=
D_{S_1}^{(w)}
+
D_{S_2}^{(w)}
+
D_{S_1S_2}^{(w)},
\qquad
D_{S_1S_2}^{(w)}
=
\sum_{i\in S_1}\sum_{j\in S_2}w_iw_jd_{ij}.
\]
For any unchanged block $U$,
\[
A_{SU}=A_{S_1U}+A_{S_2U},
\qquad
D_{SU}=D_{S_1U}+D_{S_2U}.
\]
Thus split, merge, and cut-swap proposals require recomputing only the affected
within-block and between-block factors.

\subsection{Tree-cut proposals}

At each partition update, one of three moves is proposed: split, merge, or
cut-swap. Let $p_s(K)$, $p_m(K)$, and $p_b(K)$ denote the probabilities of
choosing split, merge, and cut-swap at the current number of clusters $K$.
Boundary cases are handled by setting inadmissible move probabilities to zero,
for example $p_m(1)=0$ and $p_s(n)=0$.

\noindent\textbf{Split.}
Assume $K<n$. A split move chooses an uncut tree edge
\[
e\in E(T)\setminus B
\]
uniformly from the $n-K$ uncut edges and adds it to $B$. Cutting $e$ splits one
block $S$ into two connected blocks, denoted $S_e^-$ and $S_e^+$. The proposal
probabilities are
\[
q(x,x')=\frac{p_s(K)}{n-K},
\qquad
q(x',x)=\frac{p_m(K+1)}{K}.
\]
The local score difference is
\begin{align*}
\Delta\ell_{\mathrm{split}}
={}&
\log W_w(S_e^-)+\log W_w(S_e^+)-\log W_w(S)
+\log R(S_e^-,S_e^+) \\
&+
\sum_{U\in\rho:\,U\neq S}
\left[
\log R(S_e^-,U)+\log R(S_e^+,U)-\log R(S,U)
\right].
\end{align*}
The log Metropolis--Hastings ratio is
\[
\log r_{\mathrm{split}}
=
\log\frac{\pi_0(T,B\cup\{e\},K+1,\rho'\mid G)}
         {\pi_0(T,B,K,\rho\mid G)}
+
\Delta\ell_{\mathrm{split}}
+
\log\left\{\frac{p_m(K+1)}{K}\right\}
-
\log\left\{\frac{p_s(K)}{n-K}\right\}.
\]

\noindent\textbf{Merge.}
Assume $K>1$. A merge move chooses a cut edge $e\in B$ uniformly from the
$K-1$ cut edges and removes it from $B$. If $e$ separates two adjacent blocks
$U_e$ and $V_e$, the proposed block is
\[
W_e=U_e\cup V_e.
\]
The proposal probabilities are
\[
q(x,x')=\frac{p_m(K)}{K-1},
\qquad
q(x',x)=\frac{p_s(K-1)}{n-K+1}.
\]
The local score difference is
\begin{align*}
\Delta\ell_{\mathrm{merge}}
={}&
\log W_w(W_e)-\log W_w(U_e)-\log W_w(V_e)-\log R(U_e,V_e) \\
&+
\sum_{U\in\rho:\,U\notin\{U_e,V_e\}}
\left[
\log R(W_e,U)-\log R(U_e,U)-\log R(V_e,U)
\right].
\end{align*}
The log Metropolis--Hastings ratio is
\[
\log r_{\mathrm{merge}}
=
\log\frac{\pi_0(T,B\setminus\{e\},K-1,\rho'\mid G)}
         {\pi_0(T,B,K,\rho\mid G)}
+
\Delta\ell_{\mathrm{merge}}
+
\log\left\{\frac{p_s(K-1)}{n-K+1}\right\}
-
\log\left\{\frac{p_m(K)}{K-1}\right\}.
\]

\noindent\textbf{Cut-swap.}
Assume $K\geq 2$. A cut-swap move chooses a cut edge $e\in B$ uniformly. Let
$U$ and $V$ be the two adjacent blocks separated by $e$, and set
\[
W=U\cup V.
\]
The move temporarily removes $e$ from the cut set and then chooses a different
edge
\[
f\in E(T[W])\setminus\{e\}
\]
uniformly. Cutting $f$ produces two connected blocks, denoted $W_f^-$ and
$W_f^+$. Conditional on the selected pair of adjacent blocks, the forward and
reverse proposal probabilities are equal:
\[
q(x,x')=q(x',x)
=
\frac{p_b(K)}{(K-1)(|W|-2)}.
\]
The local score difference is
\begin{align*}
\Delta\ell_{\mathrm{swap}}
={}&
\log W_w(W_f^-)+\log W_w(W_f^+)
-\log W_w(U)-\log W_w(V) \\
&+\log R(W_f^-,W_f^+)-\log R(U,V) \\
&+
\sum_{Z\in\rho:\,Z\notin\{U,V\}}
\left[
\log R(W_f^-,Z)+\log R(W_f^+,Z)
-\log R(U,Z)-\log R(V,Z)
\right].
\end{align*}
Since $K$ is unchanged, the cut-swap proposal is symmetric and the tree-cut prior
is unchanged under the uniform cut-set prior. Therefore
\[
\log r_{\mathrm{swap}}
=
\Delta\ell_{\mathrm{swap}}.
\]
If the selected cut edge has $|W|=2$, then
$E(T[W])\setminus\{e\}$ is empty and the cut-swap move is skipped; otherwise,
$f$ is selected uniformly from $E(T[W])\setminus\{e\}$.

\noindent\textbf{Acceptance rule.}
For each of the three moves, the proposed state is accepted with probability
\[
\alpha(x,x')=\min\{1,\exp(\log r)\}.
\]

\subsection{Prior-ratio simplification under the default tree-cut prior}

The main paper uses the prior
\[
p(T\mid G)=\text{Unif}\{\text{spanning trees of }G\},
\qquad
p(K)\propto \eta^{K-1},
\]
and
\[
p(B\mid T,K)=\binom{n-1}{K-1}^{-1}.
\]
For split and merge proposals, the spanning tree $T$ is unchanged, so
$p(T\mid G)$ cancels. For a split from $K$ to $K+1$,
\[
\frac{\pi_0(x')}{\pi_0(x)}
=
\eta
\frac{\binom{n-1}{K-1}}{\binom{n-1}{K}}
=
\eta\frac{K}{n-K}.
\]
Thus the split log ratio simplifies to
\[
\log r_{\mathrm{split}}
=
\Delta\ell_{\mathrm{split}}
+
\log\eta
+
\log\frac{p_m(K+1)}{p_s(K)}.
\]
For a merge from $K$ to $K-1$,
\[
\frac{\pi_0(x')}{\pi_0(x)}
=
\eta^{-1}
\frac{\binom{n-1}{K-1}}{\binom{n-1}{K-2}}
=
\eta^{-1}\frac{n-K+1}{K-1}.
\]
Thus the merge log ratio simplifies to
\[
\log r_{\mathrm{merge}}
=
\Delta\ell_{\mathrm{merge}}
-
\log\eta
+
\log\frac{p_s(K-1)}{p_m(K)}.
\]
For cut-swap moves, $K$ and $|B|$ are unchanged, and hence
\[
\log r_{\mathrm{swap}}=\Delta\ell_{\mathrm{swap}}.
\]

\subsection{Exact tree update}

The spanning tree $T$ is an auxiliary representation of the current connected
partition $\rho$. Conditional on $\rho$ and $w$, the collapsed likelihood
depends on the tree-cut state only through $\rho$, and its full posterior conditional distribution is given by  
$P\bigl(\mathcal{T}\mid - \bigr)
=\frac{1}{\tau\bigl(M(G, \rho)\bigr)\,\prod_{h=1}^{K}\tau(G[S_h])}$,
where $\tau\bigl(M(G, \rho)$ is the number of spanning trees of the quotient graph $M(G, \rho)$ obtained
by contracting all vertices within each cluster \(S_h\) into a
single super-vertex, and $\tau(G[S_h])$ is the number of spanning trees of each subgraph. A draw from the
conditional distribution of compatible tree-cut representations has the following Gibbs step
that is always accepted.
For each block $S_h$, sample an internal spanning tree of the induced
subgraph $G[S_h]$ uniformly. Next construct the quotient multigraph whose
vertices are the blocks of $\rho$ and whose edge multiplicity between blocks
$S_h$ and $S_\ell$ is
\[
c_{h\ell}
=
\#\{(i,j)\in E_G:i\in S_h,\ j\in S_\ell\}.
\]
Sample a spanning tree $\mathcal T_\rho$ of the quotient graph with probability
proportional to
\[
\prod_{\{h,\ell\}\in E(\mathcal T_\rho)} c_{h\ell}.
\]
For each selected quotient edge $\{h,\ell\}$, choose one of the corresponding
original graph edges between $S_h$ and $S_\ell$ uniformly. Combining these
between-block edges with the internal spanning trees gives a compatible spanning
tree $T'$ from its posterior conditional distribution. The updated cut set is
\[
B'=\{e\in E(T'): e \text{ joins two different blocks of }\rho\}.
\]
The updated state is $(T',B',K,\rho,w)$ and is accepted with probability one.

\subsection{Frailty full conditional}

Fix the current partition $\rho$ and update one frailty $w_i$. Suppose
$i\in S$, where $S$ is the current block containing node $i$. Let
\[
n_S=|S|,
\qquad
m_S=\binom{n_S}{2},
\qquad
q_S=a_\lambda+\delta_w m_S.
\]
Write
\[
D_S^{(w)}
=
\sum_{u<v:\,u,v\in S}w_uw_vd_{uv}
=
C_i+w_iH_i,
\]
where
\[
C_i=
\sum_{u<v:\,u,v\in S\setminus\{i\}}w_uw_vd_{uv},
\qquad
H_i=
\sum_{j\in S:\,j\neq i}w_jd_{ij}.
\]
Only the prior $p(w_i\mid\kappa)$ and the within-block factor $W_w(S)$ depend
on $w_i$. Therefore the exact full conditional is
\[
p(w_i\mid w_{-i,S},\rho,d,\kappa)
\propto
w_i^{\kappa-1+\delta_w(n_S-1)}
\exp(-\kappa w_i)
\left(b_\lambda+C_i+w_iH_i\right)^{-q_S},
\qquad
w_i>0.
\]
For singleton blocks, $n_S=1$, $m_S=0$, $C_i=0$, and $H_i=0$, so the full
conditional reduces to the prior $\Ga(\kappa,\kappa)$.

\subsection{Log-scale frailty update}

Let
\[
\eta_i=\log w_i,
\qquad
w_i=\exp(\eta_i),
\qquad
B_i=b_\lambda+C_i.
\]
Including the Jacobian term, the log conditional density of $\eta_i$ is, up to
an additive constant,
\[
\log p(\eta_i\mid w_{-i},\rho,d,\kappa)
=
\{\kappa+\delta_w(n_S-1)\}\eta_i
-\kappa \exp(\eta_i)
-q_S\log\{B_i+H_i\exp(\eta_i)\}.
\]
A Gaussian random-walk proposal on the log scale is used:
\[
\eta_i'=\eta_i+\epsilon_i,
\qquad
\epsilon_i\sim N(0,s_i^2),
\qquad
w_i'=\exp(\eta_i').
\]
Because the proposal is symmetric in $\eta_i$, the log Metropolis--Hastings
ratio is
\[
\log r_i
=
\{\kappa+\delta_w(n_S-1)\}(\eta_i'-\eta_i)
-\kappa\{\exp(\eta_i')-\exp(\eta_i)\}
-q_S
\log
\left(
\frac{B_i+H_i\exp(\eta_i')}
     {B_i+H_i\exp(\eta_i)}
\right).
\]
The proposed frailty is accepted with probability
\[
\alpha_i=\min\{1,\exp(\log r_i)\}.
\]

The log conditional is concave in $\eta_i$. Indeed,
\[
\frac{\partial^2}{\partial \eta_i^2}
\log p(\eta_i\mid w_{-i},\rho,d,\kappa)
=
-\kappa\exp(\eta_i)
-
q_S
\frac{B_iH_i\exp(\eta_i)}
     {\{B_i+H_i\exp(\eta_i)\}^2}
<0,
\]
since $\kappa>0$, $B_i>0$, $H_i\geq 0$, and $\exp(\eta_i)>0$.


\begin{algorithm}[H]
\caption{Partially collapsed tree-cut sampler for \textsc{DISCCO}}
\label{alg:discco_sampler}
\begin{algorithmic}[1]
\STATE \textbf{Input:} distance matrix $D$, spatial graph $G$, hyperparameters
$(a_\lambda,b_\lambda,\delta_w,a_\theta,b_\theta,\delta_b,\kappa)$, tree-cut
prior parameters, and frailty proposal scales $s_1,\ldots,s_n$.
\STATE Initialize a connected partition $\rho^{(0)}$, a compatible tree-cut
representation $(T^{(0)},B^{(0)},K^{(0)})$, and positive frailties
$w^{(0)}$.
\STATE Compute the required within-block and between-block summaries.
\FOR{$t=1,\ldots,N$}
    \STATE Choose a partition move type: split, merge, or cut-swap, using
    probabilities $p_s(K)$, $p_m(K)$, and $p_b(K)$.
    \STATE Propose the corresponding local move of $(T,B,K,\rho)$.
    \STATE Recompute only the affected collapsed factors.
    \STATE Compute 
    $\log r_{\mathrm{split}}$, $\log r_{\mathrm{merge}}$, or
    $\log r_{\mathrm{swap}}$.
    \STATE Accept or reject the proposed partition move using
    $\min\{1,\exp(\log r)\}$.
    \STATE Update the tree-cut representation conditional on the accepted
    partition $\rho$.
    \FOR{$i=1,\ldots,n$}
        \STATE Compute $C_i$, $H_i$, $B_i=b_\lambda+C_i$, and
        $q_S=a_\lambda+\delta_w\binom{|S|}{2}$ for the current block
        $S\ni i$.
        \STATE Propose $\eta_i'=\eta_i+\epsilon_i$ with
        $\epsilon_i\sim N(0,s_i^2)$.
        \STATE Accept or reject using the log-scale frailty ratio $\log r_i$.
    \ENDFOR
    \STATE Store $(T^{(t)},B^{(t)},K^{(t)},\rho^{(t)},w^{(t)})$ after burn-in
    and thinning as desired.
\ENDFOR
\end{algorithmic}
\end{algorithm}


\section{Computing Resources}
\label{app:computing-resources}

All experiments were run on a MacBook Pro (2022) with an Apple M1 Pro 8-core CPU and 16 GB of RAM. No GPU acceleration was used. Reported wall-clock runtimes in the paper were obtained from CPU-based implementations on this machine. Memory usage is dominated by storage of the full pairwise distance matrix, which scales as \(O(n^2)\) in the number of spatial units.

\end{document}